\PassOptionsToPackage{table}{xcolor}
\documentclass[manuscript,screen]{acmart}
\acmJournal{TOSEM}
\setcopyright{none}
\acmDOI{}
\acmYear{}
\acmVolume{}
\acmNumber{}
\acmArticle{}

\usepackage{amsmath}
\usepackage{graphicx}
\usepackage{flafter}
\usepackage{textcomp}
\usepackage{url}
\usepackage{tikz}
\usepackage{soul}
\usepackage{comment}
\usepackage{todonotes}
\usepackage{multirow}
\usepackage{booktabs}
\usepackage{hhline}
\usepackage{siunitx}
\usepackage{subcaption}
\usepackage{algorithm}
\usepackage{algpseudocode}
\usepackage{enumitem}
\usepackage{multicol}
\usepackage[most]{tcolorbox}
\usepackage{array}
\usepackage{tabularx}
\usepackage{pifont}
\usepackage{threeparttable}
\usepackage{caption}
\usepackage{verbatim}
\usepackage{makecell}

\newcommand{\pcind}{\hspace{1.2em}}
\newcommand{\pcindtwo}{\hspace{2.4em}}
\newcommand{\pcindthree}{\hspace{3.6em}}

\definecolor{pcAccent}{HTML}{1D4ED8}
\definecolor{pcComment}{HTML}{34A853}
\definecolor{pcKeyword}{HTML}{111827}
\definecolor{pcType}{HTML}{F59E0B}

\newlength{\promptExampleGap}
\newtcolorbox{pseudobox}[1][]{%
  enhanced,
  colback=black!1,
  colframe=black!18,
  boxrule=0.4pt,
  arc=2pt,
  left=2pt,right=2pt,top=2pt,bottom=2pt,
  borderline west={1.2pt}{0pt}{pcAccent},
  sharp corners=southwest,
  sharp corners=northeast,
  before skip=0pt, after skip=0pt,
  fontupper=\normalsize\ttfamily,
  listing only,
  #1
}

\newcommand{\Fn}[1]{\textcolor{pcAccent}{\textbf{#1}}}
\newcommand{\Kw}[1]{\textcolor{pcKeyword}{\textbf{#1}}}
\newcommand{\Ty}[1]{{\textbf{#1}}}
\newcommand{\Fld}[1]{\texttt{#1}}
\newcommand{\Cmt}[1]{\textcolor{pcComment}{// #1}}
\definecolor{pcLineNo}{HTML}{9CA3AF}
\newcommand{\pcln}[1]{\textcolor{pcLineNo}{\normalsize #1}}

\newcommand{\code}[1]{\texttt{#1}}
\newcommand{\func}[1]{\texttt{#1}}
\newcommand{\var}[1]{\texttt{#1}}

\tcbuselibrary{skins,breakable}
\definecolor{promptbg}{RGB}{248,249,251}
\definecolor{promptframe}{RGB}{30,30,30}
\definecolor{prompttitle}{RGB}{50,70,120}

\newtcolorbox{promptbox}[2][]{%
  enhanced,
  colback=promptbg,
  colframe=promptframe,
  boxrule=0.5pt,
  arc=2pt,
  left=2pt,right=2pt,top=2pt,bottom=2pt,
  before skip=0pt, after skip=0pt,
  fonttitle=\bfseries\normalsize,
  coltitle=prompttitle,
  title={#2},
  attach title to upper,
  fontupper=\normalsize,
  before upper={\setlength{\parindent}{0pt}\raggedright},
  #1
}

\newcommand{\Prompt}[2]{%
  \begin{promptbox}{#1}
#2
  \end{promptbox}
}

\newcommand{\pfield}[1]{\textbf{\textsf{#1:}}\ }
\newcommand{\psep}{\par\vspace{2pt}}
\newcommand{\pmono}[1]{\texttt{#1}}
\newcommand{\pRole}[1]{\pfield{Role}#1\psep}
\newcommand{\pContext}[1]{\pfield{Context}#1\psep}
\newcommand{\pTask}[1]{\pfield{Task}#1\psep}
\newcommand{\pInput}[1]{\pfield{Input}#1\psep}
\newcommand{\pOutput}[1]{\pfield{Output}#1\psep}
\newcommand{\pConstraints}[1]{\pfield{Constraint}#1\psep}
\newcommand{\pReplace}[1]{%
  \sethlcolor{black!10}%
  \hl{\textit{#1}}%
}

\definecolor{aflnetOrange}{HTML}{FF7F0E}
\definecolor{lemurGreen}{HTML}{2CA02C}
\definecolor{semBlue}{HTML}{1F77B4}
\definecolor{afExec}{HTML}{7F1D1D}
\definecolor{afSem}{HTML}{EA580C}
\definecolor{afGram}{HTML}{EAB308}
\definecolor{soExec}{HTML}{0F172A}
\definecolor{soSem}{HTML}{0369A1}
\definecolor{soGram}{HTML}{0D9488}

\begin{document}

\title{Enhancing Protocol Fuzzing via LLM-Generated Semantic-Aware Mutators}

\author{Keyi Wang}
\email{keyiwang1010@gmail.com}
\affiliation{%
  \institution{East China Normal University}
  \city{Shanghai}
  \country{China}
}

\author{Yiutak Choi}
\email{jeremyarg9@gmail.com}
\affiliation{%
  \institution{East China Normal University}
  \city{Shanghai}
  \country{China}
}

\author{Pengda Li}
\email{pdli@stu.ecnu.edu.cn}
\affiliation{%
  \institution{East China Normal University}
  \city{Shanghai}
  \country{China}
}

\author{Chijin Zhou}
\email{tlock.chijin@gmail.com}
\affiliation{%
  \institution{East China Normal University}
  \city{Shanghai}
  \country{China}
}

\author{Haiying Sun}
\email{hysun@sei.ecnu.edu.cn}
\affiliation{%
  \department{Shanghai Key Laboratory of Trustworthy Computing}
  \institution{East China Normal University}
  \city{Shanghai}
  \country{China}
}

\author{Chengcheng Wan}
\email{ccwan@sei.ecnu.edu.cn}
\affiliation{%
  \institution{East China Normal University}
  \city{Shanghai}
  \country{China}
}
\affiliation{%
  \institution{Shanghai Innovation Institute}
  \city{Shanghai}
  \country{China}
}

\author{Ting Su}
\email{tsu@sei.ecnu.edu.cn}
\affiliation{%
  \institution{East China Normal University}
  \city{Shanghai}
  \country{China}
}

\author{Geguang Pu}
\email{ggpu@sei.ecnu.edu.cn}
\affiliation{%
  \institution{East China Normal University}
  \city{Shanghai}
  \country{China}
}

\renewcommand{\shortauthors}{Keyi Wang et al.}

\begin{abstract}
    Network protocol implementations are critical to system security but vulnerable due to their complexity.
    Mutation-based protocol fuzzing is widely used to find such flaws, but its general-purpose mutators often violate field semantics and cross-field constraints, producing invalid messages rejected early.
    This paper introduces a novel system design that enhances mutation-based protocol fuzzing with protocol-specific, semantic-aware mutators. These mutators generate more valid messages reaching deep protocol states, which general-purpose mutators could exploit to cover more edge cases.
    To this end, we use large language models (LLMs) to generate four protocol-specific components: message templates, parsers/reassemblers, semantic-aware field-level mutators, and constraint fixers, enabling raw-byte parsing, semantic field mutation, cross-field constraint repair, and message reassembly.
    To ensure the quality of these four LLM-generated components, we design three automated quality checkers: a metamorphic checker for parsers/reassemblers, a parseability checker for mutators, and a constraint-violation checker for fixers.
    We implement LEMUR-Fuzz on vanilla AFLNet and evaluate twelve implementations across ten protocols, covering all ProFuzzBench protocols with publicly available RFCs. Results show that LEMUR-Fuzz improves the average branch coverage by 6.93\% and 5.28\% over AFLNet and ChatAFL, respectively.
    LEMUR-Fuzz also discovers a previously unknown vulnerability in the latest Mosquitto, which has been confirmed and fixed by developers.
\end{abstract}

% Proposed subject classification and keywords; authors should confirm.
\begin{CCSXML}
<ccs2012>
 <concept>
  <concept_id>10011007.10011074.10011099.10011102</concept_id>
  <concept_desc>Software and its engineering~Software testing and debugging</concept_desc>
  <concept_significance>500</concept_significance>
 </concept>
</ccs2012>
\end{CCSXML}
\ccsdesc[500]{Software and its engineering~Software testing and debugging}
\keywords{Protocol fuzzing, large language models, semantic-aware mutation, software testing}
\maketitle

\section{Introduction}
\begin{figure}[t]
    \centering
    \includegraphics[width=1\linewidth]{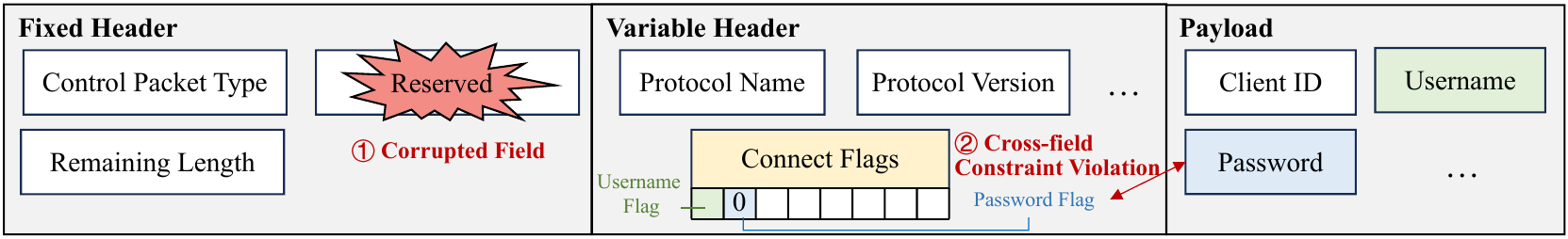}
    \caption{Two representative cases of invalid messages produced by AFLNet on MQTT CONNECT messages: \footnotesize{\textcircled{1} corruption of critical fields; and \textcircled{2} violation of cross-field constraints.
    %, such as the inconsistencies between the Password Flag in the variable header and the Password field in the payload.
    }}
    \label{fig:pic_failure}
\end{figure}

Implementing network protocols is a process of translating standardized specifications (e.g., RFCs) into executable code.
However, this process could be error-prone, and even small mistakes could lead to serious vulnerabilities. 
A well-known example is the OpenSSL Heartbleed vulnerability~\cite{cve20140160} (CVE-2014-0160), where a mistake of missing bound check in the TLS heartbeat extension led to severe sensitive data leakage. As protocols grow in complexity, manual auditing becomes impractical. It necessitates automated techniques to effectively find vulnerabilities in protocol implementations.

To this end, \emph{mutation-based} protocol fuzzing is widely adopted for vulnerability discovery~\cite{zhang2024survey, jiang2024survey, natella2022stateafl, zhang2023survey, li2022snpsfuzzer, snapfuzz, schumilo2022nyx}. 
Specifically, to perform fuzzing, such protocol fuzzers (e.g., AFLNet~\cite{aflnet}, SGFuzz~\cite{sgfuzz}, NSFuzz~\cite{nsfuzz}, ChatAFL~\cite{chatafl}) leverage \emph{mutation operators} (i.e., \emph{mutators}) to generate diverse and likely malformed message sequences from existing ones.
Such fuzzers successfully found many edge-case vulnerabilities~\cite{aflnet,sgfuzz,nsfuzz,chatafl}.
% Such fuzzers have successfully found many edge-case vulnerabilities~\cite{aflnet,sgfuzz,nsfuzz,chatafl}.

However, most of the mutators in these fuzzers are designed to perform general-purpose mutation (e.g., bit flipping, block splicing) on raw, unstructured byte streams.
They do not respect the semantics of the fields within a message or the constraints between these fields.  
% As a result, many mutated messages could be invalid and early-rejected, and are difficult to reach deep protocol states.
As a result, many mutated messages are invalid, rejected early, and unlikely to drive protocol implementations into deep states.

Indeed, our empirical analysis of MQTT v5.0~\cite{oasis-mqtt-v5} reveals this limitation. 
We use AFLNet, a state-of-the-art mutation-based protocol fuzzer, to conduct a 24-hour fuzzing campaign against Mosquitto (an implementation of MQTT).
Only a small portion (3\%) of mutated messages can be successfully processed by the MQTT server. The remaining messages are rejected during message parsing or validation.

% Figure~\ref{fig:pic_failure} illustrates two representative cases of invalid MQTT CONNECT messages, which are generated by AFLNet but rejected by the MQTT server.
Figure~\ref{fig:pic_failure} illustrates two independent ways in which AFLNet's general-purpose mutators (e.g., bit flipping) can invalidate a valid MQTT CONNECT seed.
In case~\textcircled{1}, a blind bit flip changes a reserved bit in the fixed header from 0 to 1, violating the required CONNECT format.
In case~\textcircled{2}, a separate bit flip clears the Password Flag while leaving the Password field in the payload, violating their cross-field constraint.
Both mutated messages are rejected before they can exercise deeper protocol logic.

To tackle this limitation, our \emph{key} idea is to introduce \emph{protocol-specific, semantic-aware} mutators into mutation-based protocol fuzzing.
Instead of perturbing raw bytes, these mutators operate on protocol fields and preserve field semantics and cross-field constraints.
% These new mutators could respect the
% semantics of the fields within a message and the constraints between these fields, and thus generate many likely valid messages during fuzzing.
The resulting valid messages can reach deep protocol states, from which the original general-purpose mutators can further explore edge cases.

However, manually crafting such semantic-aware mutators for a given protocol is \emph{challenging}, because it requires protocol-specific expertise and substantial effort to design mutators.
Therefore, we propose a system design that uses LLMs once before fuzzing to generate protocol-specific components, rather than invoking LLMs online to directly generate test inputs.
Specifically, we decompose this design into four LLM-generated components per protocol: (i) \emph{message templates}, which specify the structures of messages and the semantics of fields;
(ii) a \emph{parser} and a \emph{reassembler}, which lift raw byte streams (processed by existing mutation-based fuzzers) to structured messages and reassemble the
mutated messages back into byte streams;
(iii) \emph{semantic-aware mutators}, which operate at the \emph{field} level (e.g., mutating a UTF-8 encoded field while respecting its encoding), and (iv) \emph{constraint fixers}, which repair the violations of cross-field constraints after mutation by updating the content of relevant fields.
% These components are generated only in a one-time pre-fuzzing phase and reused in later fuzzing runs without further LLM calls.

A second challenge is how to ensure the quality of the four LLM-generated components. 
% Manual validation is expensive and error-prone, so we combine limited manual validation with automated correctness checkers. 
To this end, we design three types of \emph{automated} quality checkers to reveal
potential errors caused by LLM hallucinations.
Specifically, (1) for the parser and reassembler, we use the metamorphic relation $\mathrm{M}=\text{reassembler}(\text{parser}(\mathrm{M}))$ as the checker ($\mathrm{M}$ is any valid message); (2)
for semantic-aware mutators, we use the parseability property as the checker: a semantically mutated message should still be valid and could be parsed successfully;
(3) for fixers, we generate constraint-violating messages for each cross-field constraint and check whether the violations is correctly repaired.
Note that these checkers cannot guarantee the correctness of the LLM-generated components, but provide empirically effective quality checking with full automation (discussed in Section~\ref{sec:discussion}).
On the other hand, since message templates serve as the foundation of the other LLM-generated components, we do manual checking to ensure the correctness.

% To evaluate our idea,
We build a tool named LEMUR-Fuzz on top of AFLNet~\cite{aflnet}, a representative mutation-based protocol fuzzer.
The core design of LEMUR-Fuzz is to integrate semantic-aware mutation as an additional stage in AFLNet's fuzzing loop, allowing it to synergize with the original general-purpose mutators rather than replace them.
This design enables semantic-aware mutators to generate valid message sequences that drive executions into deeper protocol states, while AFLNet's original mutators continue byte-level exploration from those states.
Section~\ref{sec:implementation} details the design and implementation.

We applied LEMUR-Fuzz to fuzz twelve widely-used protocol implementations across ten protocols, covering all ProFuzzBench protocols with publicly available RFCs. Compared to state-of-the-art baselines AFLNet and ChatAFL, LEMUR-Fuzz improves branch coverage by 6.93\%, and 5.28\% on average, respectively. 
% We further integrate LLM-generated mutators and constraint fixers into Peach, a generation-based fuzzer, showing that the design is not tied to mutation-based fuzzing. This integration improves Peach's branch coverage by 5.66\% on average.
On some protocols from the protocol fuzzing benchmark ProFuzzBench~\cite{profuzzbench}, LEMUR-Fuzz can find \emph{all} the 11 bugs which are found by AFLNet, and additionally found three bugs which have not been found by AFLNet.
Moreover, within 24-hour fuzzing campaigns, LEMUR-Fuzz discovers one previously unknown vulnerability in the latest version of Mosquitto, which has been confirmed and fixed by the developers.
This paper makes the following contributions:
\begin{itemize}[leftmargin=*]
    \item At the conceptual level, we introduced semantic-aware mutators to enhance mutation-based protocol fuzzing. These mutators can generate more valid messages reaching deep protocol states, on which original general-purpose mutators could exploit to cover more edge cases. To our knowledge, we are the first to enhance protocol fuzzing by introducing protocol-specific semantic-aware mutators. 

    \item At the design level, we designed a modular system that (1) decomposes protocol knowledge into four LLM-generated components to support semantic-aware mutation; and (2) applies three types of automated quality checkers to ensure the quality of these components before fuzzing.

    %\item We automatically check LLM-generated parsers/reassemblers, mutators, and constraint fixers before fuzzing: (i) a metamorphic relation for parsers and reassemblers, (ii) post-mutation parseability for mutators, and (iii) constraint-violation tests for constraint fixers. 

    \item At the technical level, we implemented LEMUR-Fuzz to realize our idea. It has been made publicly available at \url{https://github.com/Keyi-Wang/LEMUR-Fuzz}.

    \item At the empirical level, we applied LEMUR-Fuzz to fuzz twelve protocol implementations across ten protocols, covering all ProFuzzBench protocols with publicly available RFCs. Compared to state-of-the-art baselines, LEMUR-Fuzz achieved higher branch coverage and found more unique bugs.
\end{itemize}

\section{Background}

\noindent\textbf{Concepts of Network Protocols.} 
A \emph{protocol} defines communication rules between network devices, including message format, sequencing, error handling, and timing. It is typically documented in standards such as \emph{Request for Comments} (RFCs)~\cite{rfc2026}. A \emph{message} (packet) is the basic communication unit with structured fields (e.g., header and payload). A \emph{message sequence} is an ordered sequence of messages that realizes a protocol interaction. A protocol \emph{state} records the interaction context accumulated from prior messages, such as authentication status, negotiated parameters, and session variables. A \emph{protocol implementation} is the concrete realization of this specification, including message parsing, state maintenance, and logic execution. Because protocols are often stateful, the same input message can trigger different behaviors under different states~\cite{aflnet,nsfuzz,luo2023bleem}.

\begin{figure}[t]
    \centering
    % \begin{subfigure}[t]{0.28\linewidth}
    %   \centering
    %   \includegraphics[width=\linewidth, trim=8 8 8 5, clip]{mutator/pics/03_observation_mosquitto_valid_ratio_24h.pdf}
    %   \caption{Trend in 24 hours.}
    %   \label{fig:mosq-valid-rate-24h}
    % \end{subfigure}\hfill
    % \begin{subfigure}[t]{1\linewidth}
      % \centering
      \includegraphics[width=\linewidth]{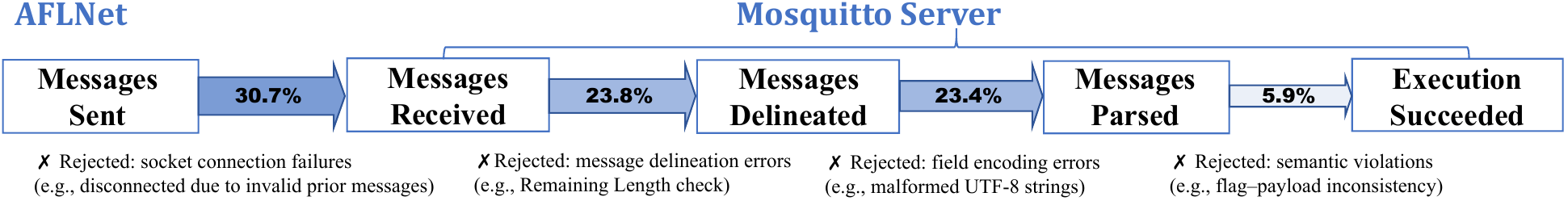}
      \caption{Message rejection breakdown across processing stages.}
      \label{fig:msg-reject-class}
    % \end{subfigure}
    % \caption{The message validity and rejection characterization c}
    % \label{fig:mosq-valid-and-reject}
  \end{figure}
  
\noindent\textbf{Mutation-based Protocol Fuzzing.}
Mutation-based protocol fuzzing discovers vulnerabilities by mutating seed inputs and sending them to the server under test (SUT). In protocol fuzzing, a seed is typically a client-to-server message sequence. AFLNet is a representative approach: it extends AFL with state guidance from server response codes for stateful protocols. For a selected target state~$s$, AFLNet partitions a seed into three segments, $M_1$, $M_2$, and $M_3$: $M_1$ drives to $s$, $M_2$ stays within $s$, and $M_3$ is the remaining suffix. AFLNet then applies havoc/splicing mutations mainly on $M_2$, and retains inputs that increase coverage or reach new states. At the mutation level, AFLNet uses general-purpose byte-level mutators (e.g., bit/byte flips, arithmetic operations, bytes insertion/deletion/repetition) and region-level mutators, but these mutators are mainly unaware of field semantics~\cite{zhangunderstanding}.

\section{Observation and Motivation}
\label{Observation}

% \hspace*{1em}
% \textbf{Observation}.
We conduct an empirical study on Mosquitto, a popular MQTT broker implementation, to investigate the limitation of general-purpose mutation in AFLNet.
We consider all mutated $M_2$ messages during AFLNet's 24-hour fuzzing campaign.
A message is regarded as \emph{execute-valid} if it is successfully processed by the server and receives a success response code. The \emph{execution success rate} is defined as the fraction of execute-valid messages among all generated messages. AFLNet achieves an extremely low success rate for mutated $M_2$ messages. Even at the earliest stage, the rate reaches only about 5.9\%, and it further deteriorates over time, stabilizing at around 3\% for the remainder of the 24-hour campaign.

% Given an initial $M_2$ message, we generate mutated messages in a single pass of the mutation stage and collect their execution outcomes.

To further understand the causes of invalid MQTT messages, we instrumented the server-side implementation to monitor the execution outcomes of all mutated $M_2$ messages based on a given seed. 
% We collected the execution outcomes of all messages mutated from the seed.
Under the effects of the original mutators (havoc and splicing), AFLNet generates a total of 7,336 mutated $M_2$ (marked as $M_2'$) messages.
As summarized in Fig.~\ref{fig:msg-reject-class}, only 30.7\% of all sent $M_2'$ messages are successfully read by the server, while the remaining messages are rejected due to socket disconnection triggered by previously sent invalid inputs. Then, only 23.8\% are successfully delineated into complete messages; the rest fail message delineation, primarily due to malformed Remaining Length encoding. After delineation, 23.4\% of messages are successfully parsed, with a small fraction being rejected due to field-level encoding errors such as malformed UTF-8 strings. More notably, only 5.9\% of all sent messages are successfully executed. The majority of parsed messages violate MQTT semantic rules or cross-field constraints, as in the two CONNECT examples discussed earlier.

% \textbf{Semantic-Aware Stage}. 
% To address the above pitfall, we design LEMUR-Fuzz to insert a \emph{semantic-aware stage} into the normal mutation process of existing mutation-based fuzzers. It includes both mutators that modify message fields while respecting intra-field semantics, and constraint fixers that repair cross-field inconsistencies after mutation.
% Figure~\ref{fig:pic_overview} illustrates the integration of semantic-aware stage into the fuzzing loop. 
% For each seed selected from the corpus, LEMUR-Fuzz not only performs the original havoc mutation, but also parses the seed into structured representation, applies semantic-aware mutators on individual fields, invokes constraint fixers to repair any cross-field violations, and finally reassembles it back to a raw byte stream.

\section{Design}

\begin{figure}[t]
  \centering
  \begin{tikzpicture}
    \node[inner sep=0] (overviewfig) {\includegraphics[width=1\linewidth]{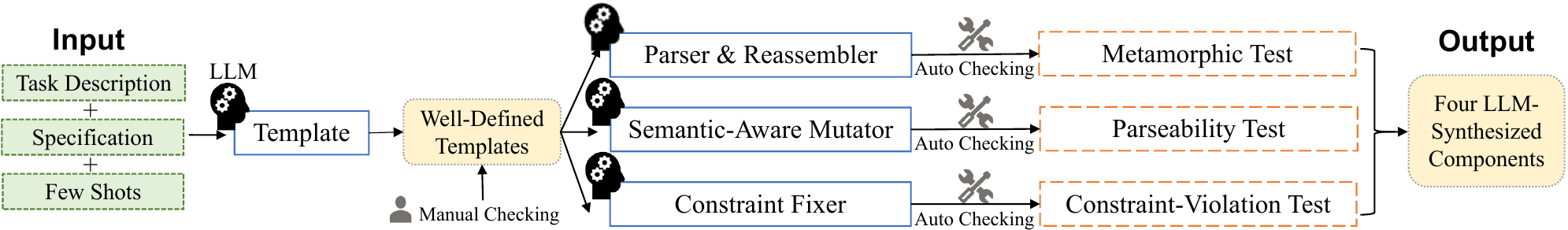}};
  \end{tikzpicture}
  \caption{The process of synthesizing four protocol-specific components via LLM and performing quality checking on these components.}
  \label{fig:llm-design_cropped}
\end{figure}

The core idea of our work is to integrate a semantic-aware mutation stage in the fuzzing loop to enhance mutation-based protocol fuzzers (detailed in Section~\ref{sec:implementation}).
It enables semantic-aware mutators to generate valid message sequences that drive executions into deep protocol states, while original mutators could exploit those states.

However, implementing the semantic-aware mutation on raw byte streams is impractical since raw bytes lack semantic information.
Therefore, we need to perform mutation on a structured view of the input: using a protocol-specific parser to lift bytes into message templates with explicit fields, applying field-level semantic-aware
mutators, and then reassembling the mutated messages back into bytes.
To enable this workflow, our approach leverages LLM to generate four protocol-specific components once per protocol from its specification: (i) message templates, (ii) a parser and a reassembler, (iii) semantic-aware field mutators, and (iv) constraint fixers for repairing cross-field inconsistencies after mutation. 

Figure~\ref{fig:llm-design_cropped} illustrates the process of generating the four components.
Each component undergoes quality checking. Message templates are manually checked before downstream component generation. Parser-reassembler pairs are checked by a metamorphic relation; mutators are checked by parseability preservation; and fixers are checked using constraint-violating tests. We detail the whole process as follows.

\subsection{Message Template Generation}
\label{cha:TemplateGen}
Message templates denote a structured, field-level representation of protocol messages.
It is the foundation for generating parsers, reassemblers, semantic-aware mutators and constraint fixers.
Specifically, message templates are generated through two-staged LLM prompts with the protocol specification.
First, we prompt the LLM by \emph{``List all the request message types of \func{<protocol\_name>}''} to obtain the list of message types of a protocol.
Second, for each message type, we ask the LLM to generate one message template that records  all the containing fields, and labels these fields as \emph{Fixed}, \emph{Optional}, or \emph{Repeatable}.
Note that one message template describes one message type, and a protocol has multiple message templates.
Based on this template, semantic-aware mutators (detailed in Section~\ref{sec:semantic-aware-mutators}) are generated, each of which only mutates the fields defined in the message template of a given message type.
% These templates provide a common field-level representation for the downstream parser, reassembler, mutators, and constraint fixers.

\begin{figure}[t]
  \centering
  \captionsetup[subfigure]{skip=6pt}

  \begin{subfigure}[t]{0.42\linewidth}
\vspace{0pt}
    \centering
    \begin{promptbox}[equal height group=template-example]{Message Template Extraction Prompt}
      \pRole{You are a protocol-aware code generator.}
      \pContext{
        The target protocol is \pReplace{MQTT}. The specification is provided and should be treated as authoritative.
      }
      \pTask{
        For the \pReplace{CONNECT} message, generate the precise C-language message structure and annotate each field using the labels listed below when applicable; fields without a matching label may be left untagged.

        \begin{itemize}
          \item \pmono{Fixed}: constant value, not mutable.
          \item \pmono{Optional}: may be present or omitted.
          \item \pmono{Repeatable}: may appear multiple times.
        \end{itemize}
      }
      \pInput{
        \pmono{[Specification]} \pmono{[Few-shots]}
      }
      \pOutput{
         Output only C code. No additional text.
      }
    \end{promptbox}

    \caption{Prompt for generating message templates}
    \label{fig:prompt_of_message_template_sub}
  \end{subfigure}%
\hspace{\promptExampleGap}%
\begin{subfigure}[t]{\dimexpr\linewidth-\promptExampleGap-0.42\linewidth\relax}
\vspace{0pt}
    \centering
    \begin{pseudobox}[fontupper=\normalsize\ttfamily,equal height group=template-example]
\begin{tabular}{@{}r@{\hspace{0.4em}}>{\raggedright\arraybackslash}p{\dimexpr\linewidth-2em\relax}@{}}
\pcln{1}  & \Kw{typedef} \Kw{struct} \{ \\
\pcln{2}  & \pcind \Ty{uint8\_t} \ \Fld{connect\_flags}; \\
\pcln{3}  & \pcind \dots \\
\pcln{4}  & \} \Ty{mqtt\_connect\_variable\_header\_t}; \\
\pcln{5}  & \Kw{typedef} \Kw{struct} \{ \\
\pcln{6} & \pcind \dots \\
\pcln{7}  & \pcind \Ty{uint8\_t} \ \Fld{will\_properties}[]; \Cmt{Optional}\\
\pcln{8}  & \pcind \Ty{uint8\_t} \ \Fld{will\_topic}[]; \Cmt{Optional}\\
\pcln{9} & \pcind \Ty{uint8\_t} \ \Fld{will\_payload}[]; \Cmt{Optional}\\
\pcln{10} & \pcind \dots \\
\pcln{11} & \} \Ty{mqtt\_connect\_payload\_t}; \\
\pcln{12} & \pcind \dots \\
\pcln{13} & \Kw{typedef} \Kw{struct} \{ \\
\pcln{14} & \pcind \Ty{mqtt\_fixed\_header\_t} \ \Fld{f\_h}; \\
\pcln{15} & \pcind \Ty{mqtt\_connect\_variable\_header\_t} \ \Fld{v\_h}; \\
\pcln{16} & \pcind \Ty{mqtt\_connect\_payload\_t} \ \Fld{payload}; \\
\pcln{17} & \} \Ty{mqtt\_connect\_packet\_t}; \\
\end{tabular}
    \end{pseudobox}

    \caption{Output Example of the generated message template for MQTT CONNECT}
    \label{fig:mqtt-connect-fields-min_sub}
  \end{subfigure}

  \caption{Prompt and output example for message template.}

  \label{fig:prompt_and_example_template}
\end{figure}

Fig.~\ref{fig:prompt_and_example_template} shows the prompt and output for the MQTT CONNECT message. The prompt provides the target protocol name, the target message type, relevant specification excerpts, and few-shot examples. The LLM then produces a template that decomposes the message into fields such as \code{connect\_flags} and \code{will\_properties}. Fields that require special handling are labeled as \emph{Fixed}, \emph{Optional}, or \emph{Repeatable}; these annotations are later used to guide semantic-aware mutation.
Before generating downstream components, two authors manually check the generated templates against the specification, focusing on whether all request message types have templates, whether fields are complete and correctly ordered, and whether the \emph{Fixed}, \emph{Optional}, and \emph{Repeatable} annotations match the specification. If either author identifies an issue, the two authors discuss it and correct the template through local edits, such as adding a missing field or adjusting the labels. In our experience, fewer than 10\% of initially generated templates require such corrections, and each correction is straightforward because it only fixes the affected fields or annotations.

\subsection{Parser and Reassembler Generation}

Given message templates and protocol specifications, both the parser and the reassembler are automatically generated by LLMs, yielding protocol-specific implementations. The parser maps raw bytes to message templates, and the reassembler serializes them back into byte streams. Generated code size varies by protocol, from about 100--400 lines for simpler protocols (e.g., FTP/SMTP) to 1,000--2,000 lines for more complex ones (e.g., RTSP).
% Concrete examples are provided in the supplementary material.

\subsection{Semantic-aware Mutators Generation}
\label{sec:semantic-aware-mutators}
% We next describe how semantic-aware mutators are designed by carefully prompting the LLM with field-level semantics to guide valid and targeted mutations.

\begin{figure}[t]
  \centering
  \captionsetup[subfigure]{skip=6pt}

\begin{subfigure}[t]{\linewidth}
\vspace{0pt}

\centering
\Prompt{Semantic-aware Mutator Generation Prompt}{
\begingroup
\setlength{\parskip}{0pt}
\setlist[enumerate]{leftmargin=*,topsep=0pt,itemsep=0pt,parsep=0pt,partopsep=0pt}
\setlength{\multicolsep}{0pt}
\setlength{\premulticols}{0pt}
\setlength{\postmulticols}{0pt}
\pTask{
For the field \pReplace{will\_topic} in the \pReplace{CONNECT} message, generate semantic-aware mutators according to the tags.
\begin{enumerate}
  \item \textbf{Fixed-value check.} If fixed, output \pmono{not mutable} and stop.
  \item \textbf{Otherwise (mutable):}
  \begin{enumerate}
    \item If optional, implement: \pmono{void add\_\pReplace{will\_topic}();} \  \pmono{void delete\_\pReplace{will\_topic}();}
    \item If repeatable, implement: \pmono{void repeat\_\pReplace{will\_topic}();}
    \item Implement \pmono{void mutate\_\pReplace{will\_topic}();} covering the following semantic dimensions if possible:
    
    \begin{enumerate}[label=\Alph*.,leftmargin=2.2em,labelwidth=1.4em,labelsep=0.4em]
      \item Canonical form
      \item Boundary values
      \item Equivalence-class alternatives
      \item Allowed bitfield / enum / range
      \item Encoding-validity preservation
      \item Padding / alignment
      \item Prefix / suffix variants
      \item Random valid mix
    \end{enumerate}
    
  \end{enumerate}
\end{enumerate}

\endgroup
}
\pInput{\pmono{[Attached message template]}}

\pOutput{ONLY output valid C code. Minimal helpers allowed. No extra text.}
}

\caption{Prompt for generating mutators for \code{will\_topic} in CONNECT message}
\label{fig:mutator_prompt}
\end{subfigure}
\par\vspace{\promptExampleGap}
\begin{subfigure}[t]{\linewidth}
\vspace{0pt}
      \centering
      \begin{pseudobox}[fontupper=\normalsize\ttfamily]
      \begin{tabular}{@{}r@{\hspace{0.4em}}>{\raggedright\arraybackslash}p{\dimexpr\linewidth-2em\relax}@{}}
      \pcln{1}  & \Kw{void} \Fn{mutate\_will\_topic}(\Ty{mqtt\_packet\_t} *pkts, \Ty{size\_t} n) \{ \\
      \pcln{2}  & \pcind \Kw{for} (\Ty{size\_t} i = 0; i < n; i++) \{ \\
      \pcln{3}  & \pcindtwo \Kw{if} (pkts[i].type != TYPE\_CONNECT) \Kw{continue}; \\
      \pcln{4}  & \pcindtwo \Ty{mqtt\_connect\_packet\_t} *c = \&pkts[i].connect; \\
      \pcln{5}  & \pcindtwo \Ty{int} m = choose\_will\_topic\_mode(); \Cmt{Canonical/Boundary/EqClass/...} \\
      \pcln{6}  & \pcindtwo \Kw{switch} (m) \{ \\
      \pcln{7}  & \pcindthree \Kw{case} \Ty{CANONICAL}: set\_will\_topic\_canonical(c); \Kw{break}; \\
      \pcln{8}  & \pcindthree \Kw{case} \Ty{BOUNDARY}: set\_will\_topic\_length(c, choose\_will\_topic\_length()); \Kw{break}; \\
      \pcln{9}  & \pcindthree \Kw{case} \Ty{EQCLASS}:  set\_will\_topic\_shape(c, choose\_will\_topic\_shape()); \Kw{break}; \\
      \pcln{10} & \pcindthree \Kw{case} \Ty{ENCODING}: set\_will\_topic\_utf8(c); \Kw{break}; \\
      \pcln{11} & \pcindthree \Kw{case} \Ty{PADDING}: add\_will\_topic\_space(c); \Kw{break}; \\
      \pcln{12} & \pcindthree \Kw{case} \Ty{PREFIX\_SUFFIX}: add\_will\_topic\_prefix\_suffix(c, choose\_will\_topic\_prefix(), choose\_will\_topic\_suffix()); \Kw{break}; \\
      \pcln{13} & \pcindthree \Kw{case} \Ty{RANDOMMIX}: random\_mix\_will\_topic(c); \Kw{break}; \\
      \pcln{14} & \pcindthree \Kw{default}: \Kw{break}; \\
      \pcln{15} & \}\ \}\ \} \\

      \end{tabular}
      \end{pseudobox}

      \caption{Output Example for the \emph{mutate} mutator of \code{will\_topic}}
      \label{fig:mutator_example}
    \end{subfigure}

  \caption{Prompt and output example for semantic-aware field-level mutators.}

  \label{fig:prompt_of_mutator}
\end{figure}

Fig.~\ref{fig:prompt_of_mutator} illustrates the prompt for generating semantic-aware mutators for field \code{will\_topic} in the MQTT CONNECT message. The prompt includes the target field (\code{will\_topic} in this example), its annotation, and the message template. LLM outputs a set of field-level mutator functions in C code. 

Specifically, field \code{will\_topic} specifies the MQTT topic on which the broker publishes the client's \code{Will} message. It is an optional field that appears only when the Will Flag is enabled, but it is neither fixed nor repeatable. Therefore, our approach generates three types of mutators for this field: \emph{add}, \emph{delete}, and \emph{mutate}. 
For brevity, Figure~\ref{fig:mutator_example} shows only the generated \emph{mutate} function (which includes many different semantic-aware mutations). The corresponding \emph{add} and \emph{delete} functions are omitted because they are straightforward.
In our approach, fixed fields are treated as immutable and excluded from mutation, optional fields have paired \emph{add} and \emph{delete} mutators, and repeatable fields are handled by \emph{repeat} mutators. 
% If a field is annotated as \emph{Fixed}, it is treated as immutable and excluded from mutation. If a field is annotated as \emph{Optional}, the LLM generates paired \emph{add} and \emph{delete} mutators. If a field is annotated as \emph{Repeatable}, the LLM generates a dedicated \emph{repeat} mutator. For mutable fields, the LLM also generates a field-specific mutation function, such as \code{mutate\_will\_topic}.

Beyond these field annotations, generating semantic-aware mutators leverages the semantic of a field. Take field \code{will\_topic} as an example, it must be a valid UTF-8 topic name (1--65,535 bytes), and allow spaces. As a Topic name, it can start with \code{\$}, but prefixes such as \code{\$SYS/} are usually reserved for broker-specific use. In practice, topic names are commonly written as slash-separated paths (e.g., \code{a/b/c}).

\begin{figure}[t]
  \centering
  \captionsetup[subfigure]{skip=6pt}

  % -------- (a) Prompts (top) --------
  \begin{subfigure}[t]{\linewidth}
\vspace{0pt}
    \centering
    \Prompt{Protocol Constraint Extraction Prompt}{
      \pTask{
      Extract \textbf{all constraints related to request (client-to-server) message formats}
      from the specification.
      }

      \pInput{\pmono{[Specification-Attachment]} \pmono{[few-shots]}}

      \pOutput{
      Output a single \pmono{.txt} file that lists all extracted constraints related to
      request message formats.
      }
      
    }

    \par\vspace{\promptExampleGap}
    \Prompt{Constraint Fixer Generation Prompt}{
      \pTask{
      Write a fixer function that enforces the following constraint.
      The fixer operates on an array of \pReplace{mqtt\_packet\_t} and must \textbf{repair violations in place},
      modifying the message fields so that the constraint is satisfied after fixing.
      }

      \pConstraints{\pReplace{If the Will Flag is set to 1, the Will Properties, Will Topic, and Will Payload fields MUST be present in the Payload.}}

      \pInput{\pmono{[Attached message template]}}

      \pOutput{
      Output ONLY C code that implements the fixer function.
      Do not include explanations or additional text.
      }
    }

    \caption{Prompt for generating a fixer for Will Flag}
    \label{fig:prompt_constraint_and_fixer}
  \end{subfigure}
\par\vspace{\promptExampleGap}
\begin{subfigure}[t]{\linewidth}
\vspace{0pt}
      \centering
      \begin{pseudobox}[fontupper=\normalsize\ttfamily]
      \begin{tabular}{@{}r@{\hspace{0.4em}}>{\raggedright\arraybackslash}p{\dimexpr\linewidth-2em\relax}@{}}
      \pcln{1}  & \Ty{void} \Fn{fix\_mqtt\_will\_consistency}(\Ty{mqtt\_packet\_t} *pkts, \Ty{int} n) \{ \\
      \pcln{2}  & \pcind \Kw{for} (\Ty{int} i = 0; i < n; i++) \{ \\
      \pcln{3}  & \pcindtwo \Kw{if} (pkts[i].type != \Ty{TYPE\_CONNECT}) \Kw{continue}; \\
      \pcln{4}  & \pcindtwo \Ty{mqtt\_connect\_packet\_t} *c = \&(pkts[i].connect); \\
      \pcln{5}  & \pcindtwo \Ty{uint8\_t} flags = c->variable\_header.connect\_flags; \\
      \pcln{6}  & \pcindtwo \Ty{bool} willFlag = \Ty{has\_will\_flag}(flags); \hspace{0.6em} \Cmt{The Will Flag corresponds to bit 2 of the CONNECT flags.} \\
      \pcln{7}  & \pcindtwo \Kw{if} (!willFlag) \Kw{continue}; \\
      \pcln{8} & \pcindtwo \Kw{if} (!\Ty{has\_will\_properties}(c)) \Ty{ensure\_will\_properties}(c); \\
      \pcln{9} & \pcindtwo \Kw{if} (!\Ty{has\_will\_topic}(c))      \Ty{ensure\_will\_topic}(c); \\
      \pcln{10} & \pcindtwo \Kw{if} (!\Ty{has\_will\_message}(c))    \Ty{ensure\_will\_message}(c); \\
      \pcln{11} & \} \} \\
      \end{tabular}
      \end{pseudobox}

      \caption{Output Example for the fixer of Will Flag.}
      \label{fig:fixer_will_consistency_cstyle}
    \end{subfigure}

  \caption{Prompt and output example for constraint fixer generation.}

  \label{fig:constraint_fixer_pipeline}
\end{figure}

To this end, we design eight complementary semantic dimensions to prompt LLM to systematically generate se\-mantic-aware mutators. 
In the eight semantic dimensions, \emph{canonical forms} generate specification-recommended values to stay on the protocol's ``happy path'' (e.g., \code{sensors/temp} for  field \code{will\_topic}), while \emph{boundary values} target off-by-one and limit-related errors (e.g., empty string, or maximum valid length). \emph{Equivalence-class alternatives} explore semantically valid but structurally distinct patterns (e.g., \code{a/b} vs.\ \code{a/b/c} for \code{will\_topic}). \emph{Encoding-validity preservation} ensures compliance with strict UTF-8 requirements, which is essential for \code{will\_topic}. \emph{Padding/alignment} stresses parser robustness without altering semantics (e.g., padding bytes or boundary-aligned lengths). \emph{Prefix/suffix variants} introduce specification-allowed extensions (e.g., reserved prefixes \code{\$SYS/} or suffix tags like \code{/debug}). Dimensions such as \emph{allowed bitfield/enum/range} apply to other fields (e.g., \code{connect\_flags}) where explicit value sets are defined. Finally, \emph{random valid mixing} combines multiple semantic dimensions while preserving overall validity, increasing diversity and avoiding collapse to a single mutation pattern. Guided by these semantics-aware prompts, our approach generates a tailored set of mutators for each protocol field, enabling targeted and efficient mutations.

% Consistency between separately generated templates and mutators is ensured by using the same message template in mutator generation and fuzzing. First, during mutator generation, the message template is included in the prompt. Each template defines a set of \code{field\_name}s for one message type, and the LLM generates mutator functions only for these fields, using names such as \code{add\_<field\_name>}, \code{delete\_<field\_name>}, \code{repeat\_<field\_name>}, and \code{mutate\_<field\_name>}. Second, during fuzzing, once a packet is parsed as a specific message type, the semantic-aware mutation stage mutates it only with the mutators generated for that message type. 
% Thus, the template and the mutators are inherently consistent.
% This preserves structural consistency and keeps typed C structures stable, while cross-type exploration is left to seed diversity and AFLNet's original havoc/splicing stages.

\subsection{Constraint Fixers Generation}
\label{sec:constraint_fixers}
While some specifications (e.g., MQTT) provide appendices summarizing protocol constraints, in most cases such constraints are scattered throughout the document. Therefore, we leverage the LLM’s ability to reason over specification text to extract message-format constraints directly.

As shown in Figure~\ref{fig:prompt_constraint_and_fixer}, we provide the LLM with relevant specification excerpts and a predefined output format to guide constraint extraction.
Then, for each extracted constraint, we instruct the LLM to generate a corresponding \emph{fixer}: a function that repairs mutated seeds to respect the constraint.
The lower part of Figure~\ref{fig:prompt_constraint_and_fixer} illustrates the constraint-to-fixer generation process through an example. The constraint specifies that when a feature-indicating flag is enabled in the message, all fields required by that feature must be present.
In this example, the generated fixer (Figure~\ref{fig:fixer_will_consistency_cstyle}) enforces this consistency rule by scanning a sequence of messages, identifying those that enable the feature, and checking whether the corresponding required fields are missing. If any required field is absent, the fixer explicitly inserts a valid placeholder rather than discarding the message. This repair-oriented behavior preserves message validity and enables further protocol processing and mutation.

\subsection{Quality Checking of LLM-Generated Components}
\label{sec:component_correctness_checks}
In our approach, the core components (parsers/reassembers, mutators and fixers) are generated by LLMs, which naturally raises a fundamental question: How reliable is LLM-generated code in practice?
To address this, we design a quality-checking pipeline for these components. For each component, we execute test cases until either (i) the line coverage reaches 100\% under our harness, or (ii) a fixed testing budget is exhausted when full coverage is impractical. Specifically, we design three automated quality checkers to improve the quality of LLM-generated components.
Although these checkers cannot formally guarantee the correctness of these components, they are effective in mitigating LLM-induced errors in practice (discussed in Section~\ref{sec:discussion}).

% \vspace{2pt}
\noindent\textbf{Metamorphic Test for Parser and Reassembler.}
We design a metamorphic checker (inspired by metamorphic testing~\cite{chen2020metamorphic}) to test LLM-generated parsers and reassemblers. For any syntactically valid message sequence, parsing followed by reassembling should preserve the original byte stream without loss or modification. This yields the following metamorphic relation for any valid message:
\[
\forall \mathrm{Msg},\mathrm{Msg} = \text{reassembler}(\text{parser}(\mathrm{Msg}))
\]
We use this relation and available valid messages to expose potential errors (e.g., missing field handling or oversimplified control flow) in the generated parsers and reassemblers. 
In practice, we use publicly available seed messages of the given protocol and the valid messages generated by LLM.

% \vspace{2pt}
\noindent\textbf{Parseability Test for Mutators.}
Semantic-aware mutators are expected to have a parseability property that the mutated messages should remain syntactically valid and should still be successfully parsed, thereby avoiding premature rejection by the server. We therefore design a \emph{parseability} checker:
\[
\forall \mathrm{Msg},
% \mathrm{parseable}(\mathrm{Msg}) \Rightarrow
\mathrm{parser}\!\left(\mathrm{reassembler}\!\left(\mathrm{mutate}\!\left(\mathrm{parser}(\mathrm{Msg})\right)\right)\right) = \top,
\]

Here, $\top$ means that the checking terminates normally. The test inputs used for mutator checking are the same as those used for parser and reassembler checking, ensuring that mutators are exercised by diverse valid messages. 
In practice, given a test input (i.e., a  valid message), we parse it, apply the mutator under test on it, reassemble the mutated message into a byte stream, and parse it again. Because the implementation of these mutators is algorithmically random, we test each mutator repeatedly with 500 trials on each valid message. A mutator passes the checker only if every one of these 500 trials terminates normally.

% \vspace{2pt}
\noindent\textbf{Constraint-violation Test for Fixers.}
Checking LLM-generated fixers requires assessing whether the generated fixer implementations are consistent with their corresponding constraint descriptions in natural language. This problem amounts to checking the consistency between a piece of specification and the corresponding code.
It is difficult to automate such a task with full guarantee.
To provide practical assurance, we design a constraint-violation checker to test the fixers. Given a piece of constraint description in natural language, we prompt the LLM to generate multiple representative test inputs that deliberately violate the specified constraint. For example, to violate the MQTT Will Flag's constraint, the generated inputs would set \var{Will Flag} as \code{1} but omit one of the fields \var{Will Properties}, \var{Will Topic}, and \var{Will Payload}.
These constraint-violating inputs are then used to exercise the corresponding LLM-generated fixer. After applying the fixer, the checker asserts whether all the required Will-related fields are present to satisfy the constraint. If the fixer successfully repairs such violations and produces constraint-satisfying messages, the fixer is qualified.

\begin{figure}[t]
    \centering
    \begin{tikzpicture}
      \node[inner sep=0] (loopfig) {\includegraphics[width=1\linewidth]{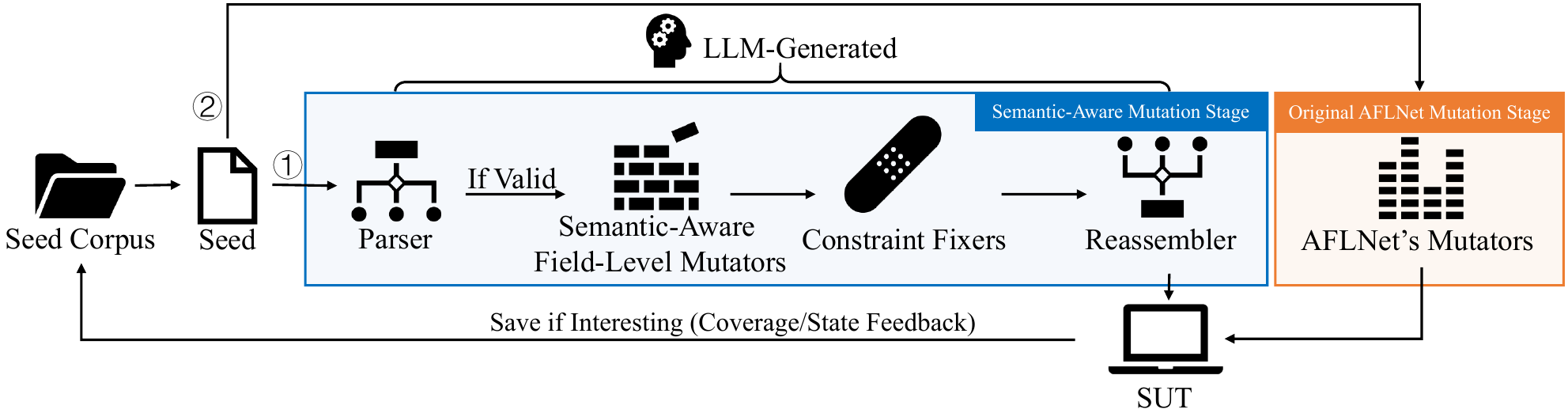}};
    \end{tikzpicture}
    \caption{An illustration of LEMUR-Fuzz integrating semantic-aware mutation. For each seed, it attempts to stage \textcircled{1} to parse the seed. If parsing succeeds, it applies semantic-aware mutators, constraint fixers, and reassembler; otherwise, it skips this stage. The same seed then enters stage \textcircled{2},  original AFLNet mutation stage.}
    \label{fig:pic_overview}
\end{figure}

\begin{algorithm}[t]
\captionsetup{font=normalsize}
\caption{Mutation augmented with semantic-aware mutators}
\label{alg:mutation-loop}
\normalsize
\algrenewcommand\alglinenumber[1]{\normalsize #1:}
\begin{algorithmic}[1]
\Procedure{AugmentedMutationStage}{$seed,\ protocol$}
  \State $pkts \gets \Call{Parser}{seed,\ protocol}$ \Comment{Parse a byte string into structured messages}
  \If{$\Call{Count}{pkts} = 0$}
    \State $\Call{OriginalMutationStage}{seed}$ \Comment{Fallback to original mutation}
    \State \Return
  \EndIf

  \For{$t \gets 1$ \textbf{to} $max\_iters$}
    \State $pkts_{init} \gets \Call{Clone}{pkts}$ \Comment{Avoid accumulating side effects across iterations}
    \State $r \gets \Call{RandomInt}{1,\ R_{max}}$

    \Statex \Comment{Apply semantic-aware field-level mutation}
    \State $pkts_{mutated} \gets \Call{SemanticAwareMutation}{pkts_{init},\ protocol,\ r}$
    \State $pkts_{fixed} \gets \Call{FixConstraints}{pkts_{mutated},\ protocol}$ \Comment{Repair protocol constraints}
    \State $seed' \gets \Call{Reassembler}{pkts_{fixed},\ protocol}$ \Comment{Serialize structured messages to bytes}
    \State \Call{Test}{$seed'$} \Comment{Send to SUT}
  \EndFor
  \State $\Call{OriginalMutationStage}{seed}$ \Comment{Fallback to original mutation}
  \State \Return
\EndProcedure
\end{algorithmic}
% \vspace{-0.3em}
\end{algorithm}
\noindent\textbf{Actions for quality checking failures.}
When the preceding quality checking fails for a parser/reassembler or fixer, we inspect the failing inputs, manually fix the code issues, and run the corresponding checkers again. This repair process requires lightweight manual effort (detailed in Section~\ref{sec:discussion}), because the failures are often tied to a small part of the generated code, such as a missing field or an incorrect length calculation.
For mutators, we discard those that fail the parseability test. We choose this strategy because a parseability failure does not readily pinpoint what is wrong with a mutator: the mutated value may break the target field's format or violate a constraint involving other fields. Diagnosing and repairing a failing mutator would require substantial manual effort. Removing these failing mutators has limited impact on fuzzing, because each field is typically associated with several mutators.

\subsection{Implementation}
\label{sec:implementation}
%TODO：figure7 and algo1. How to combine.?
% We realize the proposed method by integrating the generated protocol-aware components into a standard protocol fuzzing framework, where a semantic-aware mutation stage (Algorithm~\ref{alg:mutation-loop}) extends the original workflow with structured message processing while preserving compatibility with byte-level mutation.
We implement our approach by integrating LLM-generated protocol-specific components into AFLNet, namely LEMUR-Fuzz (Algorithm~\ref{alg:mutation-loop}, Figure~\ref{fig:pic_overview}). The key design is to add a semantic-aware mutation stage alongside AFLNet’s original byte-level mutation stage within the same fuzzing loop.
The two stages are complementary: semantic-aware mutation improves \emph{exploration} by preserving message validity to reach deep states, while original general-purpose mutation improves \emph{exploitation} by perturbing inputs around discovered states.

For each selected seed, LEMUR-Fuzz first parses raw bytes into structured messages (Alg.~\ref{alg:mutation-loop}, line~2). If parsing fails, it directly falls back to AFLNet’s original mutation stage (Alg.~\ref{alg:mutation-loop}, lines~3-6). If parsing succeeds, it iteratively performs semantic-aware field-level mutation, constraint fixing, reassembly, and execution, and then invokes AFLNet’s original havoc stage on the same seed (Alg.~\ref{alg:mutation-loop}, lines~7-15). This design makes semantic-aware mutation a complementary component rather than a replacement for generic mutators.

\section{Evaluations}
\label{sec:evaluations}
% \subsection{Experimental Design}
% To comprehensively evaluate the effectiveness, design rationale, and generality of LEMUR-Fuzz, 
In this section, we evaluate LEMUR-Fuzz and aim to answer the following research questions (RQs):
\begin{itemize}
  \item \textbf{RQ1:} Does LEMUR-Fuzz outperform state-of-the-art mutation-based protocol fuzzers?
  \item \textbf{RQ2:} Are LEMUR-Fuzz's semantic-aware mutation and AFLNet's original mutation complementary?
  % Why is augmenting a traditional mutation-based fuzzing pipeline with a semantic-aware stage more effective than replacing existing mutation mechanisms?
  \item \textbf{RQ3:} How much do semantic-aware mutators and constraint fixers contribute to LEMUR-Fuzz?
  \item \textbf{RQ4:} Can LEMUR-Fuzz's semantic-aware mutation improve message validity during protocol fuzzing?
  % \item \textbf{RQ5:} Can semantic-aware mutation generalize to generation-based fuzzers such as Peach?
\end{itemize}

\subsection{Experiment Setup}

\noindent\textbf{Benchmarks.}
Our benchmark comprises twelve protocol implementations spanning ten protocol specifications: MQTT (Mosquitto and NanoMQ), RTSP (live555), SIP (Kamailio), FTP (ProFTPD and Pure-FTPd), SMTP (Exim), DTLS (TinyDTLS), DNS (dnsmasq), DICOM (DCMTK), TLS (OpenSSL), and SSH (OpenSSH). We selected subjects commonly tested by prior protocol fuzzing studies~\cite{aflnet,chatafl,sgfuzz,psg,nsfuzz,luo2024parallel,profuzzbench} based on three dimensions: (1) different message formats, including \emph{text protocols with flexible headers} (RTSP/SIP), \emph{command-driven protocols} (FTP/SMTP), and \emph{binary protocols} (MQTT/DTLS/DNS/DICOM/TLS/SSH); (2) different scales of protocol specifications, ranging from 32 to 318 pages; and (3) different application domains, including IoT messaging (MQTT), multimedia streaming (RTSP), VoIP signaling (SIP), file transfer (FTP), email transport (SMTP), secure datagram communication (DTLS), name-service resolution (DNS), medical image networking (DICOM), secure transport (TLS), and remote shell access (SSH). The evaluated protocols cover all protocols with publicly available RFCs in ProFuzzBench. DAAP, which is in ProfuzzBench, but is excluded in our benchmark because it is a proprietary protocol without a publicly available RFC specification. The exact source commits and generated components are recorded in Table~\ref{tab:servers_and_components}.

% \textbf{Benchmarks.}
% To evaluate effectiveness and generalizability, we select representative open-source implementations of widely used protocols. Table~\ref{tab:servers_and_components} summarizes the targets and the scale of LLM-generated components. We choose six protocols, MQTT~\cite{oasis-mqtt-v5}, RTSP~\cite{rfc7826}, SIP~\cite{rfc3261}, FTP~\cite{rfc959}, SMTP~\cite{rfc5321}, and DTLS~\cite{rfc6347}, to span diverse characteristics: (i) message format complexity, covering both text-based protocols with flexible headers (RTSP/SIP), command-driven protocols with simple per-message syntax but diverse message types (SMTP/FTP) and compact binary protocols with strict encodings (MQTT/DTLS); (ii) RFC specification scale, spanning short documents (32 pages) to long documents (318 pages); and (iii) application domains, including IoT messaging, streaming, VoIP signaling, file transfer, email transport, and secure datagram communication. This diversity supports a robust assessment of LEMUR-Fuzz.
% These protocols are also widely used in prior protocol fuzzing research~\cite{aflnet, chatafl, sgfuzz, psg, nsfuzz, luo2024parallel}.

\begin{table}[t]
\centering
\caption{Statistics of protocol servers under test (latest versions) and the LLM-synthesized components per protocol.}
\label{tab:servers_and_components}

\normalsize
\setlength{\tabcolsep}{3pt}
\begin{threeparttable}
\begin{tabular*}{\linewidth}{@{\extracolsep{\fill}}lrrllrrrr@{}}
\toprule
\multicolumn{3}{c}{\textbf{Protocol specifications}} &
\multicolumn{2}{c}{\textbf{Servers under test}} &
\multicolumn{4}{c}{\textbf{Generated components}} \\
\cmidrule(lr){1-3}\cmidrule(lr){4-5}\cmidrule(l){6-9}
\makecell[c]{\bfseries Protocol} &
\makecell[c]{\bfseries \#Msg\\\bfseries Types} &
\makecell[c]{\bfseries \#RFC\\\bfseries Pages} &
\makecell[c]{\bfseries Impl.} &
\makecell[c]{\bfseries Commit} &
\makecell[c]{\bfseries \#Msg\\\bfseries Tmpl.} &
\makecell[c]{\bfseries Parser/\\\bfseries Reasm (LOC)} &
\makecell[c]{\bfseries \#Mut.} &
\makecell[c]{\bfseries \#Fix.} \\
\midrule

\multirow{2}{*}{MQTT~\cite{oasis-mqtt-v5}}
 & \multirow{2}{*}{11} & \multirow{2}{*}{137}
 & Mosquitto & \texttt{c6c6850}
 & \multirow{2}{*}{11} & \multirow{2}{*}{533 / 349} & \multirow{2}{*}{285} & \multirow{2}{*}{40} \\
 &  &
 & NanoMQ    & \texttt{f22bae7}
 &  &  &  &  \\
\arrayrulecolor{gray!45}\cmidrule(lr){1-9}\arrayrulecolor{black}

RTSP~\cite{rfc7826} & 11 & 318 &
live555   & \texttt{f4a4e8f}
& 11 & 1127 / 2005 & 242 & 7  \\
\arrayrulecolor{gray!45}\cmidrule(lr){1-9}\arrayrulecolor{black}

SIP~\cite{rfc3261}  & 6  & 269 &
Kamailio  & \texttt{bf24361}
& 6  & 893 / 592  & 536 & 17 \\
\arrayrulecolor{gray!45}\cmidrule(lr){1-9}\arrayrulecolor{black}

\multirow{2}{*}{FTP~\cite{rfc959}}
 & \multirow{2}{*}{33} & \multirow{2}{*}{69}
 & ProFTPD   & \texttt{76b7a76}
 & \multirow{2}{*}{33} & \multirow{2}{*}{360 / 240} & \multirow{2}{*}{543} & \multirow{2}{*}{11} \\
 &  &
 & Pure-FTPd & \texttt{3818577}
 &  &  &  &  \\
\arrayrulecolor{gray!45}\cmidrule(lr){1-9}\arrayrulecolor{black}

SMTP~\cite{rfc5321} & 16 & 95 &
Exim      & \texttt{6441327}
& 16 & 442 / 167 & 246 & 9  \\
\arrayrulecolor{gray!45}\cmidrule(lr){1-9}\arrayrulecolor{black}

DTLS~\cite{rfc6347} & 5  & 32 &
TinyDTLS  & \texttt{87cd057}
& 5  & 565 / 426 & 451 & 13 \\
\arrayrulecolor{gray!45}\cmidrule(lr){1-9}\arrayrulecolor{black}
DNS~\cite{rfc1034,rfc1035,rfc6891} & 6 & 126 & dnsmasq & \texttt{03fd0c1} & 6 & 523 / 360 & 74 & 29 \\
\arrayrulecolor{gray!45}\cmidrule(lr){1-9}\arrayrulecolor{black}
DICOM~\cite{dicom-ps3-7,dicom-ps3-8} & 8 & 74 & DCMTK & \texttt{ea1f2e8} & 8 & 776 / 595 & 89 & 13 \\
\arrayrulecolor{gray!45}\cmidrule(lr){1-9}\arrayrulecolor{black}
TLS~\cite{rfc5246} & 5 & 104 & OpenSSL & \texttt{122db7e} & 5 & 445 / 391 & 36 & 7 \\
\arrayrulecolor{gray!45}\cmidrule(lr){1-9}\arrayrulecolor{black}
SSH~\cite{rfc4251,rfc4252,rfc4253,rfc4254} & 8 & 103 & OpenSSH & \texttt{73c1399} & 8 & 604 / 535 & 80 & 8 \\
\bottomrule
\end{tabular*}
\begin{tablenotes}[flushleft]
\footnotesize
\item[] \emph{MsgTypes}: number of message categories represented by the generated templates; \emph{RFC Pages}: page count of the protocol RFC/standard; \emph{Impl.}: server implementation under test; \emph{MsgTemplates}: number of generated message templates; \emph{Parser/Reasm}: lines of code of the parser and reassembler; \emph{Mut.}: number of registered semantic mutators; \emph{Fix.}: number of public constraint-fixer entry points.
\end{tablenotes}

\end{threeparttable}
\end{table}

\noindent\textbf{LLM-generated components for LEMUR-Fuzz.}
As summarized in Table~\ref{tab:servers_and_components}, we use each protocol specification to guide LLM-based generation of protocol-specific components (message templates, parser/reassembler, mutators, and fixers), which are then quality-checked, refined, and integrated into AFLNet. We generate one top-level template per reported message category (5--33 per protocol); nested DICOM, TLS, and SSH message IDs remain fields within their enclosing templates. Parsers/reassemblers range from hundreds to 2,005 LOC, with 36--543 field-level semantic mutators and 7--40 constraint fixers per protocol. For test adequacy, our checking harness uses an explicit stopping rule: parsers/reassemblers/fixers are tested until 100\% line coverage on reachable code, and mutator checking is capped at 500 tests per mutator.

\begin{table}[t]
\centering
\caption{Branch coverage of LEMUR-Fuzz compared with two baselines over 24 hours (10 runs averaged).}
\label{tab:branch_cmp_lemur_vs_baselines}
\normalsize
\setlength{\tabcolsep}{0.5pt}
\begin{threeparttable}
\begin{tabular*}{\linewidth}{@{\extracolsep{\fill}}lrrrrrrrrr@{}}
\toprule
\multirow{2}{*}{\textbf{Target}} &
\multirow{2}{*}{\makecell{\textbf{LEMUR-}\\\textbf{Fuzz}}} &
\multicolumn{4}{c}{\textbf{vs. AFLNet}} & \multicolumn{4}{c}{\textbf{vs. ChatAFL}} \\
\cmidrule(lr){3-6}\cmidrule(l){7-10}
& & \textbf{Branch} & \textbf{Improv.} & \textbf{$p$-value} & \textbf{$A_{12}$ (95\% CI)}
& \textbf{Branch} & \textbf{Improv.} & \textbf{$p$-value} & \textbf{$A_{12}$ (95\% CI)} \\
\midrule
Mosquitto & 2423.2 & 2166.2 & 11.86\% & $<0.05$ & $1.00$\,[1.00,1.00] & N/A & N/A & N/A & N/A \\
NanoMQ & 3141.8 & 2779.2 & 13.05\% & $<0.05$ & $1.00$\,[1.00,1.00] & N/A & N/A & N/A & N/A \\
Kamailio & 10814.8 & 9886.2 & 9.39\% & $<0.05$ & $1.00$\,[1.00,1.00] & 9957.4 & 8.61\% & $<0.05$ & $1.00$\,[1.00,1.00] \\
live555 & 3145.0 & 2895.8 & 8.61\% & $<0.05$ & $1.00$\,[1.00,1.00] & 3000.9 & 4.80\% & $<0.05$ & $1.00$\,[1.00,1.00] \\
ProFTPD & 5316.1 & 5097.8 & 4.28\% & $<0.05$ & $0.91$\,[0.72,1.00] & 5298.7 & 0.33\% & $0.52$ & $0.59$\,[0.32,0.83] \\
Pure-FTPd & 1275.9 & 1184.7 & 7.70\% & $<0.05$ & $0.89$\,[0.70,1.00] & 1133.1 & 12.60\% & $<0.05$ & $1.00$\,[1.00,1.00] \\
Exim & 3938.3 & 3907.7 & 0.78\% & $0.39$ & $0.62$\,[0.35,0.85] & 3936.7 & 0.04\% & $<0.05$ & $0.87$\,[0.68,1.00] \\
TinyDTLS & 958.4 & 932.7 & 2.76\% & $<0.05$ & $1.00$\,[1.00,1.00] & N/A & N/A & N/A & N/A \\
dnsmasq & 1479.7 & 1335.5 & 10.80\% & $<0.05$ & $1.00$\,[1.00,1.00] & N/A & N/A & N/A & N/A \\
DCMTK & 7173.5 & 6495.5 & 10.44\% & $<0.05$ & $0.98$\,[0.91,1.00] & N/A & N/A & N/A & N/A \\
OpenSSL & 13547.0 & 13469.1 & 0.58\% & $<0.05$ & $0.78$\,[0.54,0.97] & N/A & N/A & N/A & N/A \\
OpenSSH & 2193.4 & 2131.6 & 2.90\% & $<0.05$ & $1.00$\,[1.00,1.00] & N/A & N/A & N/A & N/A \\
\midrule
\textbf{Average} & 4617.3 & 4356.8 & \textbf{6.93\%} & -- & -- & - & \textbf{5.28\%} & -- & -- \\
\bottomrule
\end{tabular*}
\begin{tablenotes}[flushleft]
\footnotesize
\item[] The AFLNet average improvement is computed as the arithmetic mean across all twelve targets; the ChatAFL average uses the targets with available ChatAFL results. N/A indicates that no ChatAFL comparison is available for that target. The $A_{12}$ values are computed from the preserved ten-run branch-hit samples using the Vargha--Delaney definition; intervals are percentile 95\% bootstrap CIs.
\end{tablenotes}
\end{threeparttable}
\end{table}

\begin{figure}[t]
  \centering
  \begin{minipage}{0.03\textwidth}
    \centering
    \rotatebox{90}{Branches covered}
  \end{minipage}%
  \begin{minipage}{0.97\textwidth}
    \centering
  
  \begin{subfigure}[t]{0.245\textwidth}
    \centering
    \includegraphics[width=\linewidth]{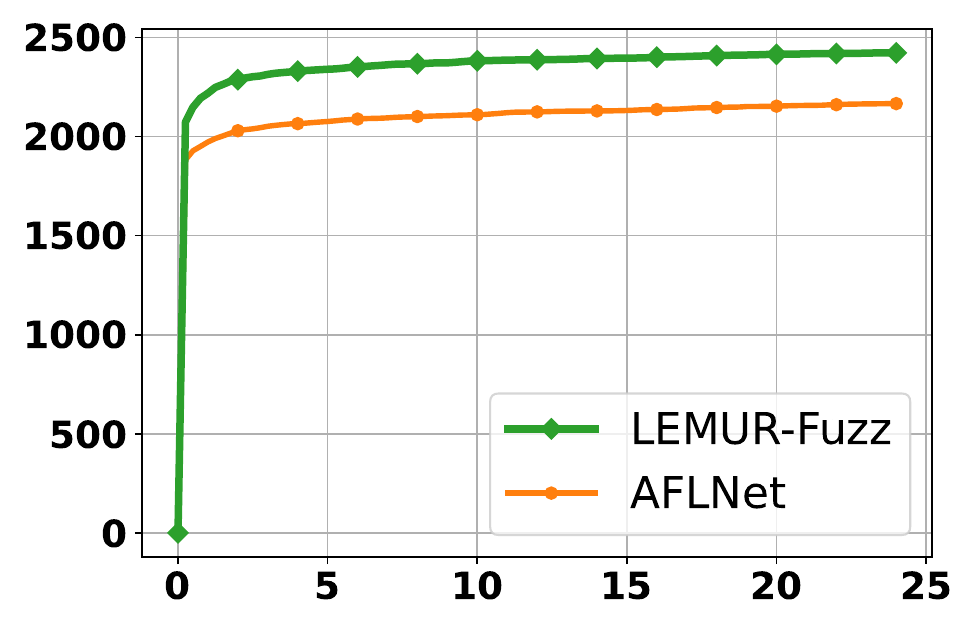}
    \caption{Mosquitto}
  \end{subfigure}\hfill
  \begin{subfigure}[t]{0.245\textwidth}
    \centering
    \includegraphics[width=\linewidth]{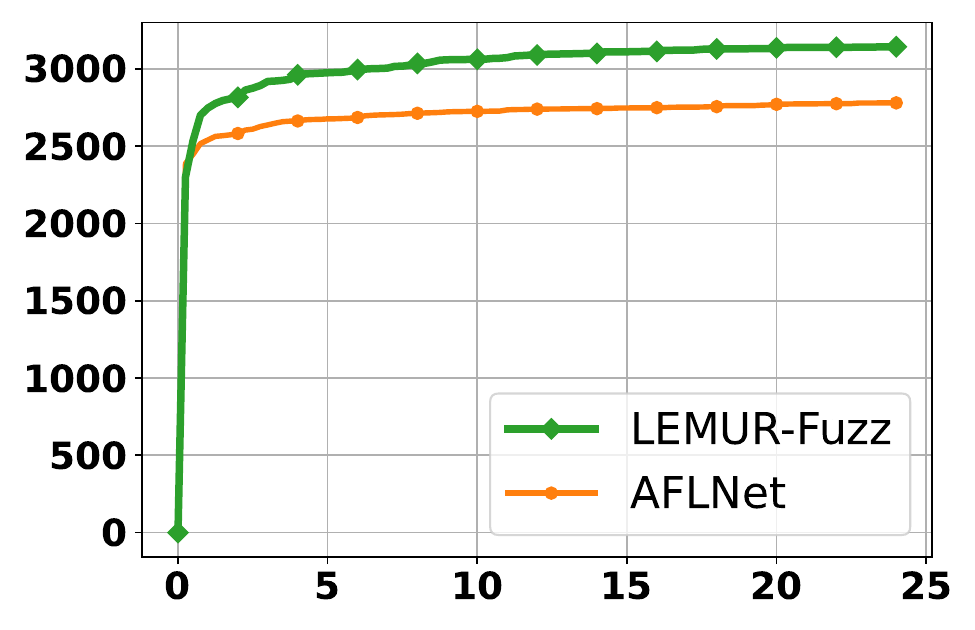}
    \caption{NanoMQ}
  \end{subfigure}\hfill
  \begin{subfigure}[t]{0.245\textwidth}
    \centering
    \includegraphics[width=\linewidth]{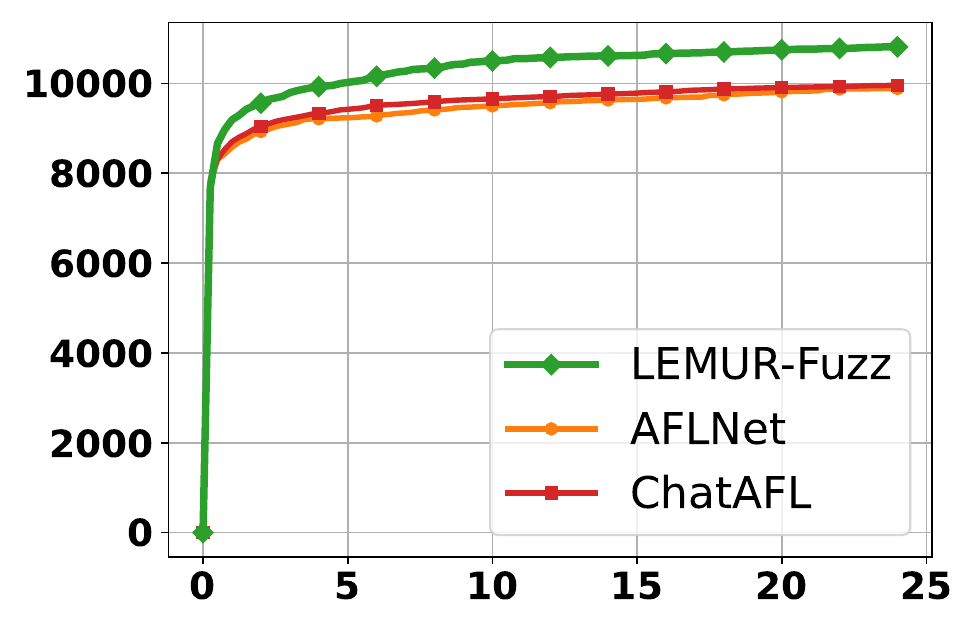}
    \caption{Kamailio}
  \end{subfigure}\hfill
  \begin{subfigure}[t]{0.245\textwidth}
    \centering
    \includegraphics[width=\linewidth]{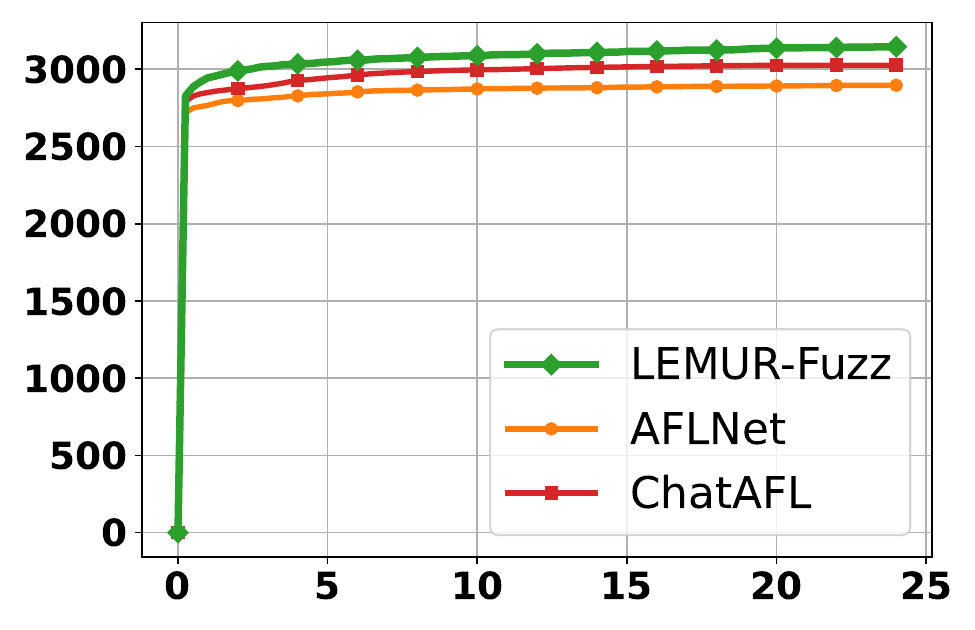}
    \caption{live555}
  \end{subfigure}

  \begin{subfigure}[t]{0.245\textwidth}
    \centering
    \includegraphics[width=\linewidth]{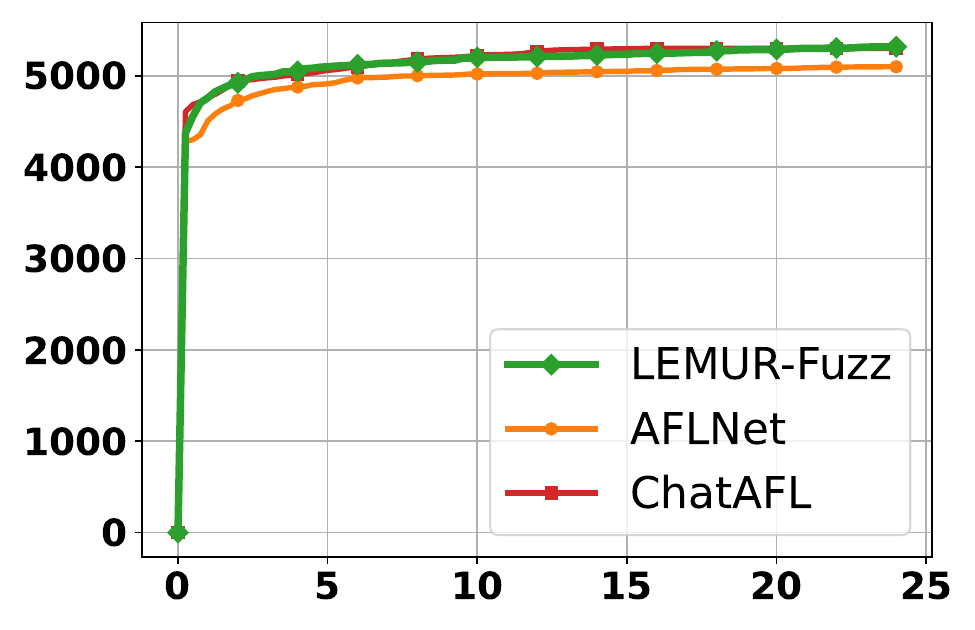}
    \caption{ProFTPD}
  \end{subfigure}\hfill
  \begin{subfigure}[t]{0.245\textwidth}
    \centering
    \includegraphics[width=\linewidth]{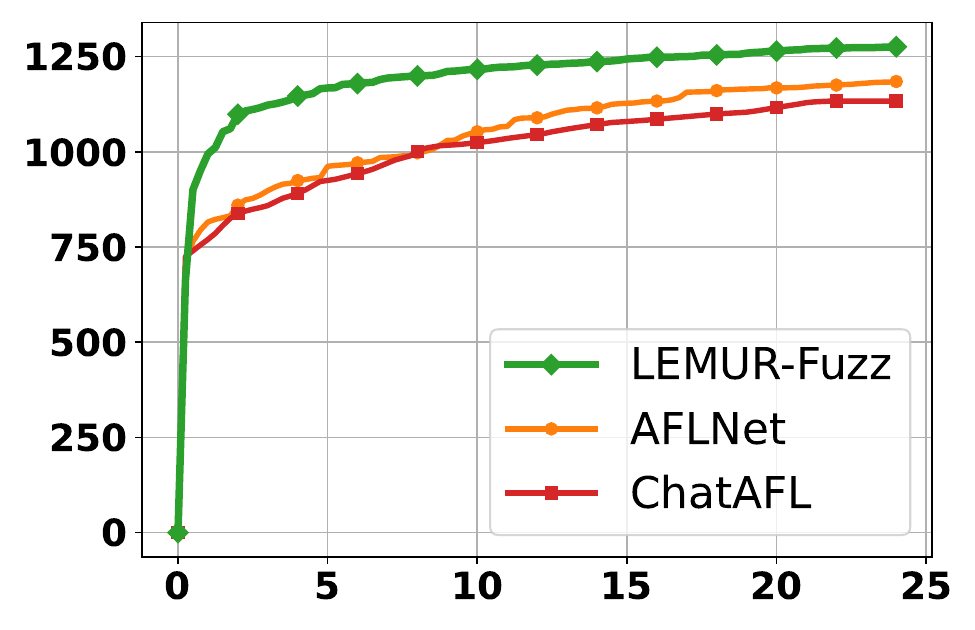}
    \caption{Pure-FTPd}
  \end{subfigure}\hfill
  \begin{subfigure}[t]{0.245\textwidth}
    \centering
    \includegraphics[width=\linewidth]{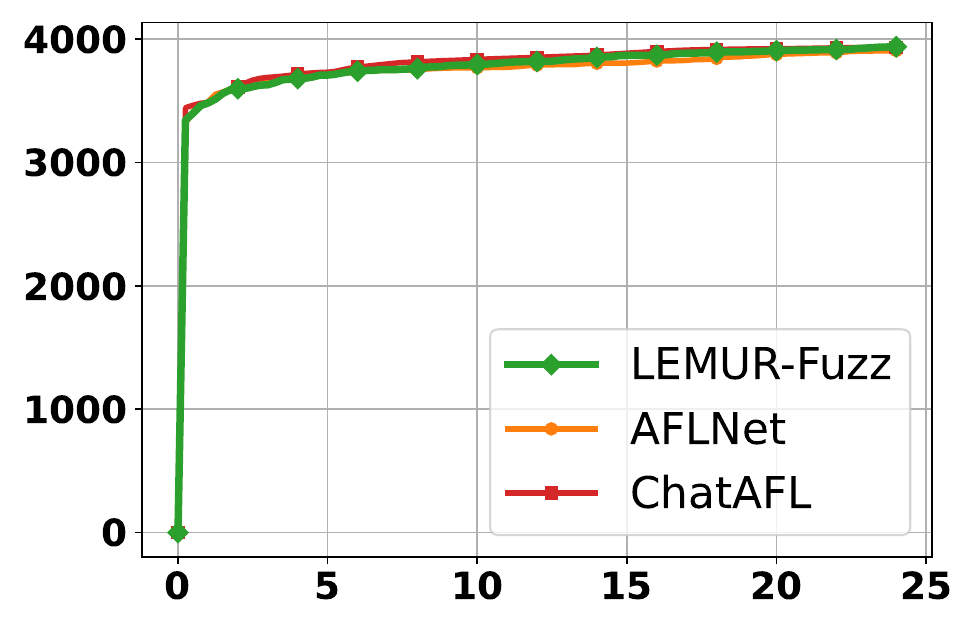}
    \caption{Exim}
  \end{subfigure}\hfill
  \begin{subfigure}[t]{0.245\textwidth}
    \centering
    \includegraphics[width=\linewidth]{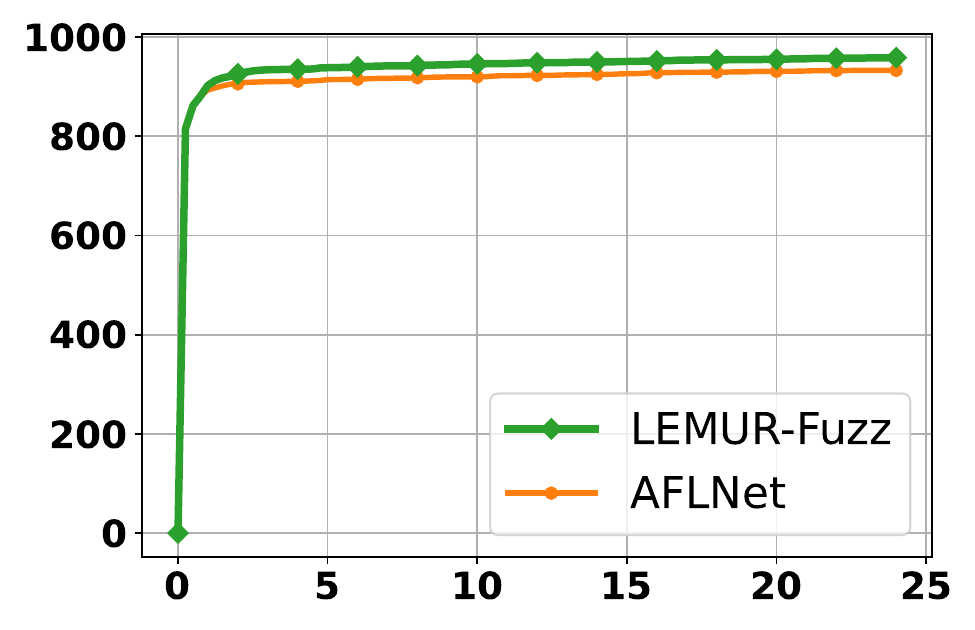}
    \caption{TinyDTLS}
  \end{subfigure}

  \begin{subfigure}[t]{0.245\textwidth}
    \centering
    \includegraphics[width=\linewidth]{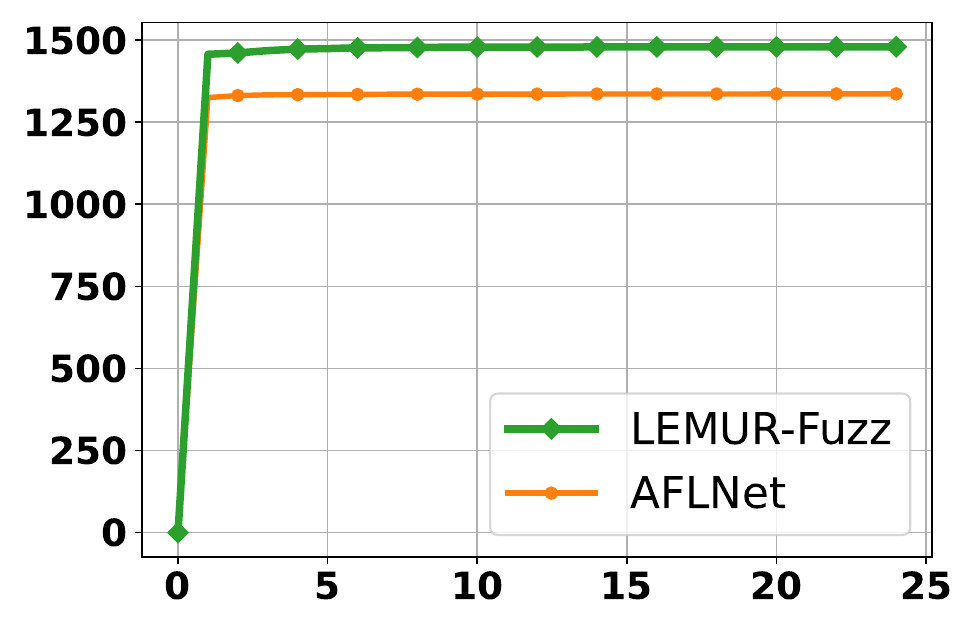}
    \caption{dnsmasq}
  \end{subfigure}\hfill
  \begin{subfigure}[t]{0.245\textwidth}
    \centering
    \includegraphics[width=\linewidth]{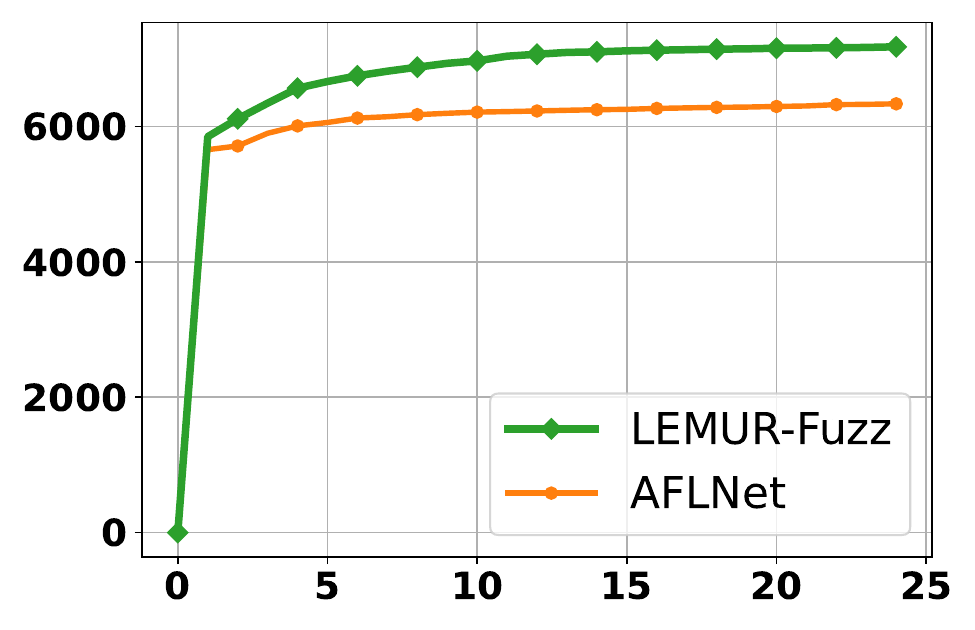}
    \caption{DCMTK}
  \end{subfigure}\hfill
  \begin{subfigure}[t]{0.245\textwidth}
    \centering
    \includegraphics[width=\linewidth]{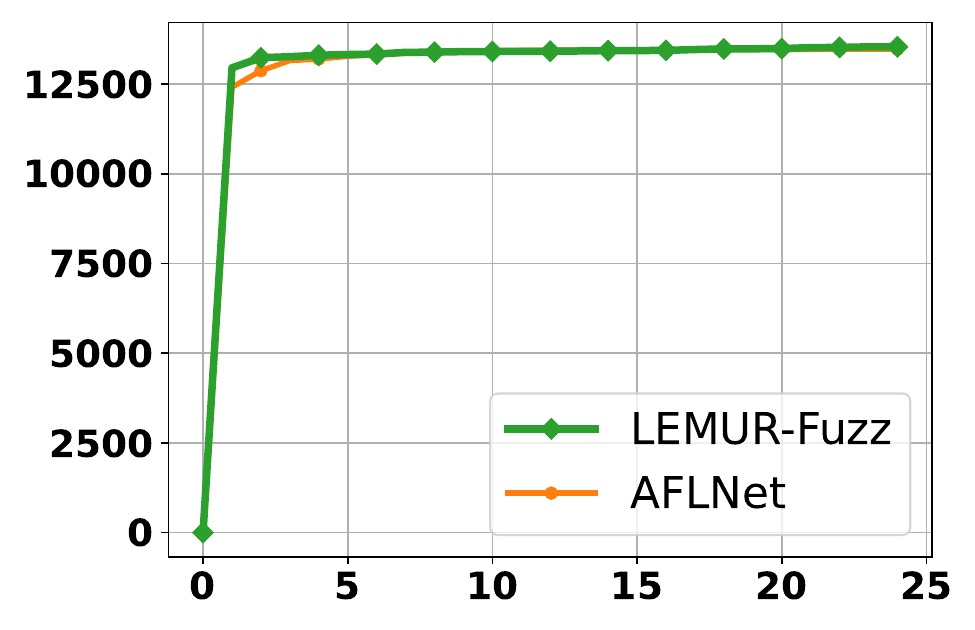}
    \caption{OpenSSL}
  \end{subfigure}\hfill
  \begin{subfigure}[t]{0.245\textwidth}
    \centering
    \includegraphics[width=\linewidth]{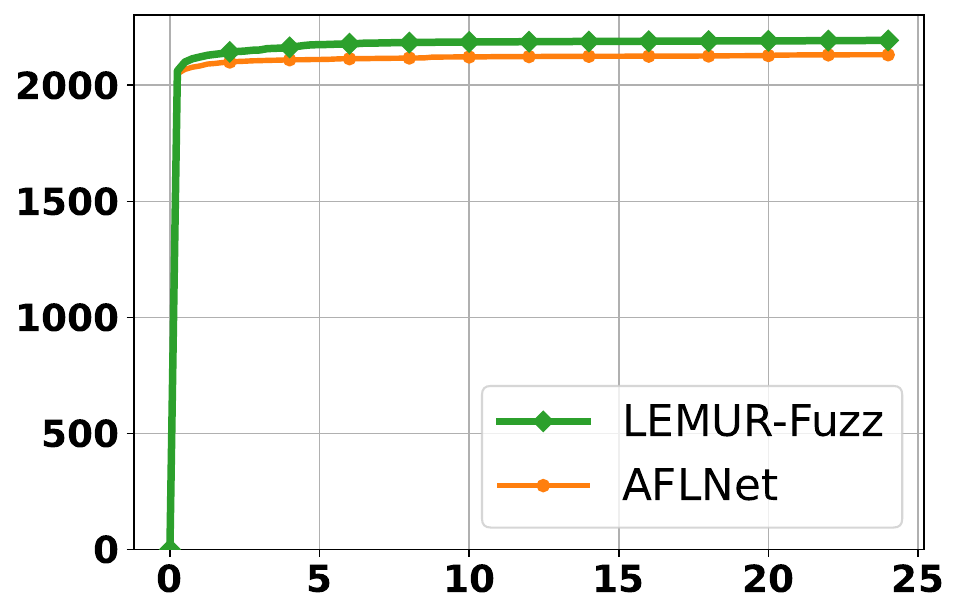}
    \caption{OpenSSH}
  \end{subfigure}

  \end{minipage}

  {Time (h)}
  \caption{Unique branch coverage achieved by different fuzzers within 24 hours over 10 independent runs.}
  \label{fig:total_cov}
  
\end{figure}

\noindent\textbf{Baselines.}
We compare LEMUR-Fuzz with AFLNet and ChatAFL across the twelve implementations. ChatAFL results are available only for the text-oriented targets for which the corresponding baseline campaign was conducted; the remaining entries are marked N/A. The $A_{12}$ values are computed from the preserved ten-run branch-hit samples.

% RQ5 selects Peach~\cite{peach}, a state-of-the-art generation-based fuzzer, to evaluate the generalizability of our semantic-aware mutation (Section~\ref{sec:rq5_peach}).

% \textbf{Settings.}
% We follow the standard evaluation methodology for fuzzing~\cite{klees2018evaluating}.
% For each configuration, we conduct 10 independent fuzzing runs, each lasting 24 hours. We also report p-value under the Mann-Whitney U test~\cite{mann1947test} to show the statistical significance.
% All experiments are conducted on a machine equipped with an AMD Ryzen Threadripper PRO 3995WX 64-Cores CPU, running Ubuntu 20.04 with 128GB of main memory. 
% % Unless otherwise stated, we follow the default runtime parameters used in prior work, so that differences arise only from the fuzzing technique itself. 
% For each target, the initial seed corpus is collected from  ProFuzzBench~\cite{profuzzbench} and AFLNet tutorials~\cite{aflnet-tutorials}; 
% All fuzzers use the same initial seed corpus.
% All software environments are kept consistent across all fuzzers to ensure a fair comparison.

\noindent\textbf{Settings.}
We follow standard fuzzing evaluation practice~\cite{klees2018evaluating}: each configuration is run 10 times for 24 hours; we report statistical significance using the Mann--Whitney U test~\cite{mann1947test} and effect size using the Vargha--Delaney $A_{12}$ statistic~\cite{vargha2000critique}. Experiments are conducted on an AMD Ryzen Threadripper PRO 3995WX (64 cores) running Ubuntu 20.04 with 128\,GB RAM. For each target, seeds are collected from ProFuzzBench~\cite{profuzzbench} and AFLNet tutorials~\cite{aflnet-tutorials}; all fuzzers use the same seed corpus and software environment for fairness.

\noindent\textbf{LLM Configuration.}
We configure LEMUR-Fuzz with the GPT-5 model to generate core components in a one-time pre-fuzzing phase. For fair comparison, we also configure ChatAFL to use the same GPT-5 model.

\noindent\textbf{LLM cost.}
Generating the components used by the twelve-implementation benchmark requires about 17M input tokens and 0.5M output tokens, costing roughly \$26 in total under the current GPT-5 API pricing. This is a one-time offline cost; MQTT and FTP implementations share protocol-level components. For comparison, a single 24-hour ChatAFL run with the same GPT-5 model costs about \$12.

\subsection{RQ1: Branch Coverage and Bug Finding}

We evaluate the effectiveness of LEMUR-Fuzz using branch coverage and bug-finding capability, comparing it against AFLNet and ChatAFL. Notably, ChatAFL is designed to test text-based protocols and is applied only to our text-based targets (RTSP/SIP/SMTP/FTP). We mark it as N/A for binary protocols (MQTT/DTLS/DNS/DICOM/TLS/SSH), where it is not applicable~\cite{chatafl}.

\subsubsection{Branch Coverage}
Table~\ref{tab:branch_cmp_lemur_vs_baselines} summarizes the average branch coverage of LEMUR-Fuzz, AFLNet, and ChatAFL over 24 hours, aggregated across ten independent runs for each of the twelve implementations.
The branch coverage is collected by \texttt{gcov}~\cite{gcov}.
The reported $p$-values are computed using the Mann--Whitney U test.
% For each target, it reports the absolute branch coverage of LEMUR-Fuzz, the corresponding coverage of each baseline, the relative improvement, and statistical significance (p-value~\cite{mann1947test}), with missing entries indicating unsupported targets. 
LEMUR-Fuzz exceeds AFLNet on all twelve implementations. Averaged over all twelve implementations, LEMUR-Fuzz reaches 4,617.3 branches compared with 4,356.8 for AFLNet, an improvement of 6.93\%.
The largest gains occur for NanoMQ, Mosquitto, dnsmasq, DCMTK, and Kamailio. These targets expose structured or state-dependent message fields---including MQTT properties and length encodings, DNS sections and name representations, DICOM UL/DIMSE nesting, and SIP header or transaction variants---where semantic mutation and constraint repair can preserve validity while exploring combinations that byte-level mutations often discard. The smallest gains occur for OpenSSL (+0.58\%), Exim (+0.78\%), TinyDTLS (+2.76\%), and OpenSSH (+2.90\%). These protocols either have compact message formats or have already been extensively covered by AFLNet at the parsing stage, while deeper coverage is gated by cryptographic handshakes, authentication, encryption, or strict state transitions; consequently, many semantic mutations either remain on paths AFLNet already reaches or are rejected before entering new code.
Figure~\ref{fig:total_cov} shows the branch-coverage timelines for the complete twelve-implementation benchmark, and Table~\ref{tab:branch_cmp_lemur_vs_baselines} reports the corresponding aggregate values.
% in stateful protocol fuzzing.

\subsubsection{Bug Finding}
We evaluate LEMUR-Fuzz against two baselines using the latest benchmark implementations. Under this setting, only LEMUR-Fuzz identifies one previously unknown bug in Mosquitto~\texttt{c6c6850}. This is because the evaluated targets have undergone extensive testing and maintenance, making new bug discovery in their latest versions inherently challenging.

To further assess the bug-finding ability of LEMUR-Fuzz, we additionally conduct experiments on earlier protocol versions that are known to be vulnerable, provided by ProFuzzBench~\cite{profuzzbench} and the AFLNet tutorials~\cite{aflnet-tutorials}. These versions provide reproducible crash oracles for the bug-finding subset of the twelve-implementation benchmark.
Over a 24-hour fuzzing campaign with 10 independent runs per target on all available vulnerable protocol versions, the three fuzzers uncovered reproducible bugs in three protocol implementations (Mosquitto~\texttt{2665705}, Live555~\texttt{ceeb4f4}, and TinyDTLS~\texttt{683cce2}) and did not find any bugs in other protocols or versions.
We summarize the results in Table~\ref{tab:bug_summary_transposed}.

% \hl{todo: change to GPT-5}

The three fuzzers collectively uncover 14 unique reproducible bugs, where uniqueness is determined by AddressSanitizer stack-trace deduplication~\cite{Serebryany2012ASan}. LEMUR-Fuzz finds all 14 bugs, while AFLNet finds 11/14 and misses TinyDTLS-7, TinyDTLS-8, and Mosquitto-3 across all 10 runs. ChatAFL is limited to text-based protocols and thus cannot be applied to TinyDTLS and Mosquitto. It also fails to discover two live555 bugs in all 10 runs. Notably, Mosquitto-3 is a previously unknown vulnerability first discovered by LEMUR-Fuzz and later confirmed and fixed by Mosquitto developers.

% Overall, the three fuzzers collectively uncover 14 unique reproducible bugs. Bug uniqueness is determined by deduplicating crashes based on AddressSanitizer~\cite{Serebryany2012ASan} stack traces. LEMUR-Fuzz successfully discovers all 14 bugs, whereas AFLNet finds only 11 out of 14, failing to trigger three bugs across all 10 runs (TinyDTLS-7, TinyDTLS-8, and Mosquitto-3). ChatAFL is limited to text-based protocols and consistently failed to discover the two bugs in live555 across all 10 independent runs.

% Notably, the bug labeled mosquitto-3 corresponds to a previously unknown vulnerability that was first discovered by LEMUR-Fuzz. This issue was subsequently confirmed and fixed by Mosquitto developers, demonstrating LEMUR-Fuzz's ability to expose real-world vulnerabilities.
\begin{table}[t]
\centering
\normalsize
\setlength{\tabcolsep}{3pt}
\caption{Reproducible unique bugs triggered within 24 hours over 10 independent runs.}
\label{tab:bug_summary_transposed}

\begin{threeparttable}
\begin{tabular}{l l ccc}
\toprule
\textbf{Bug (Target-Instance)} &
\textbf{Bug Type} &
\textbf{LEMUR-Fuzz} &
\textbf{AFLNet} &
\textbf{ChatAFL} \\
\midrule
TinyDTLS-1      & global-buffer-overflow & 10/10 & 10/10 & N/A \\
TinyDTLS-2      & global-buffer-overflow & 10/10 & 9/10  & N/A \\
TinyDTLS-3      & heap-use-after-free    & 8/10  & 10/10 & N/A \\
TinyDTLS-4      & global-buffer-overflow & 1/10  & 2/10  & N/A \\
TinyDTLS-5      & global-buffer-overflow & 7/10  & 8/10  & N/A \\
TinyDTLS-6      & global-buffer-overflow & 1/10  & 1/10  & N/A \\
TinyDTLS-7      & global-buffer-overflow & 2/10  & 0/10  & N/A \\
TinyDTLS-8      & assertion failed       & 1/10  & 0/10  & N/A \\
\midrule
live555-1       & stack-buffer-overflow  & 2/10  & 1/10  & 0/10  \\
live555-2       & SEGV                   & 10/10 & 5/10  & 9/10  \\
live555-3       & SEGV                   & 2/10  & 3/10  & 0/10 \\
\midrule
Mosquitto-1     & SEGV                   & 10/10 & 10/10 & N/A \\
Mosquitto-2     & DEADLYSIGNAL           & 9/10  & 9/10  & N/A \\
Mosquitto-3\tnote{$\dagger$} & SEGV       & 5/10  & 0/10  & N/A \\
\bottomrule
\end{tabular}

% \vspace{0.2em}
\begin{tablenotes}[flushleft]
% \tiny
\item Each cell reports the number of runs (out of 10) in which the corresponding bug was triggered at least once. [$\dagger$] denotes the newly discovered vulnerability. N/A indicates that the fuzzer is not applicable on the target.
\end{tablenotes}
\end{threeparttable}
\end{table}

\subsection{RQ2: Design Choice}
\label{sec:design_choice}
\begin{figure}[t]
  \centering
  \setlength{\tabcolsep}{1.5pt} 
  \renewcommand{\arraystretch}{1.0}

  \begin{tabular}{@{}cccc@{}}
    \subcaptionbox{Mosquitto\label{fig:venn-mosq}}
      {\includegraphics[width=0.24\linewidth, trim=8 8 8 30.6, clip]{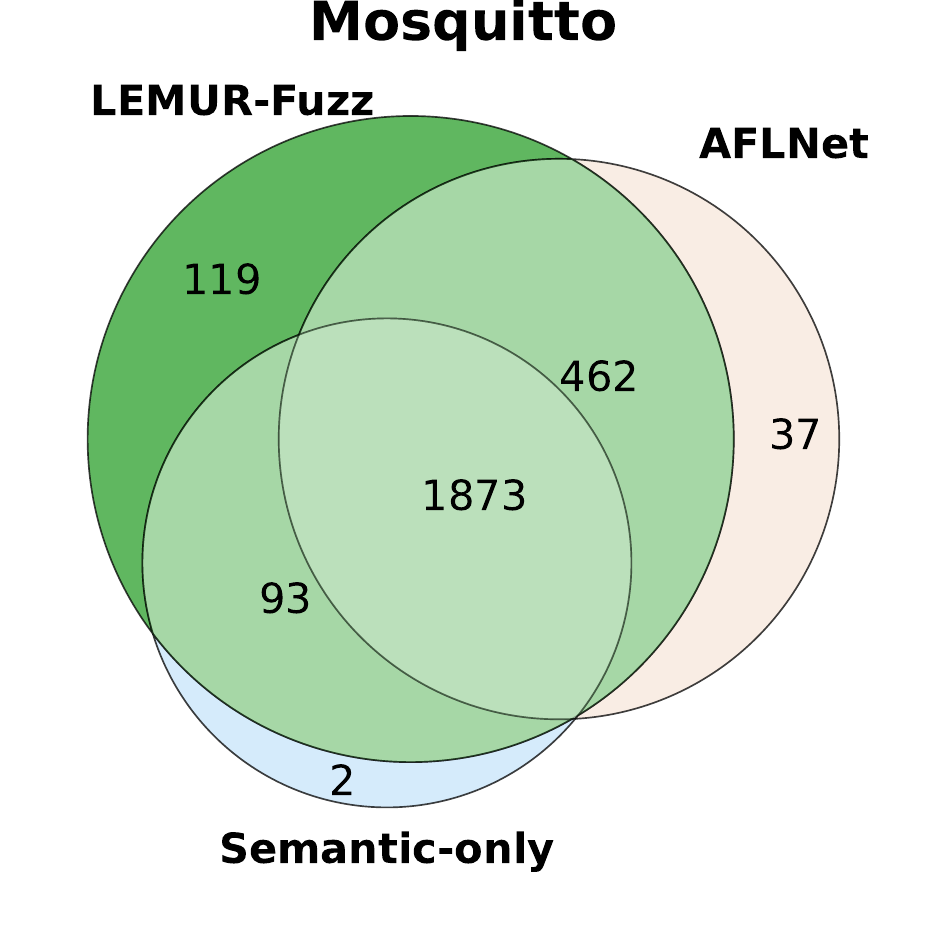}} &
    \subcaptionbox{NanoMQ\label{fig:venn-nanomq}}
      {\includegraphics[width=0.24\linewidth, trim=8 8 8 35.2, clip]{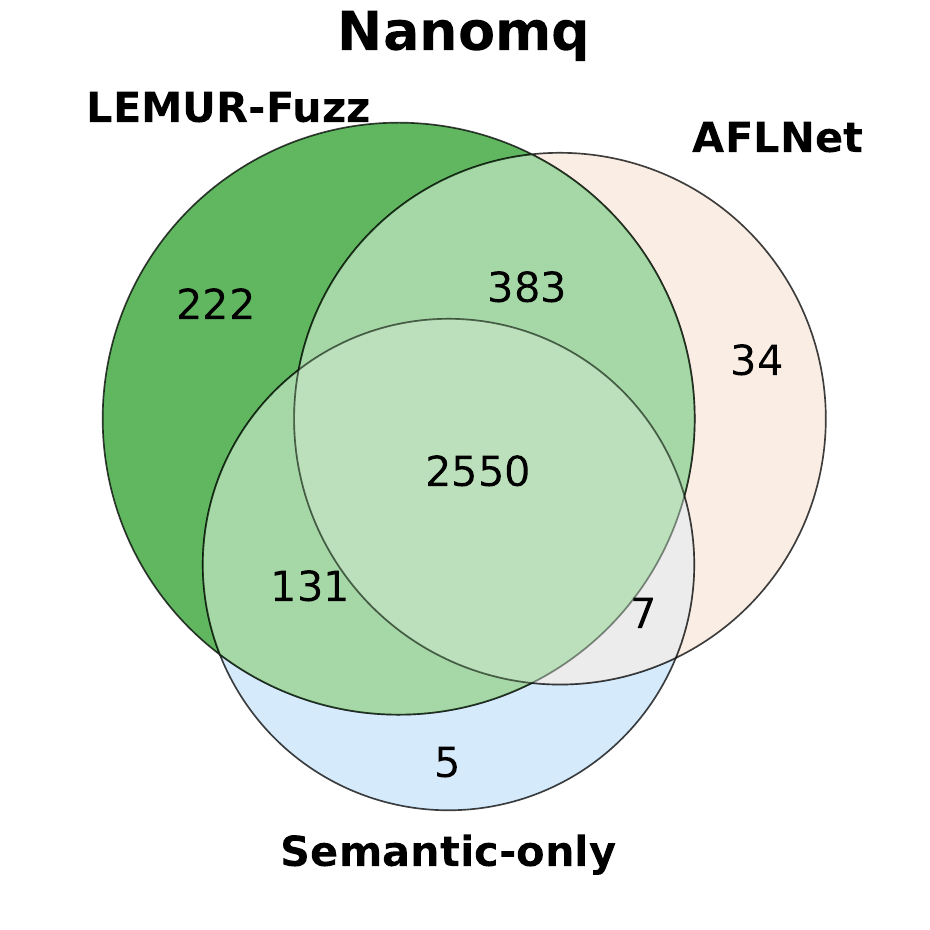}} &
    \subcaptionbox{Kamailio\label{fig:venn-kamailio}}
      {\includegraphics[width=0.24\linewidth, trim=8 8 8 33.1, clip]{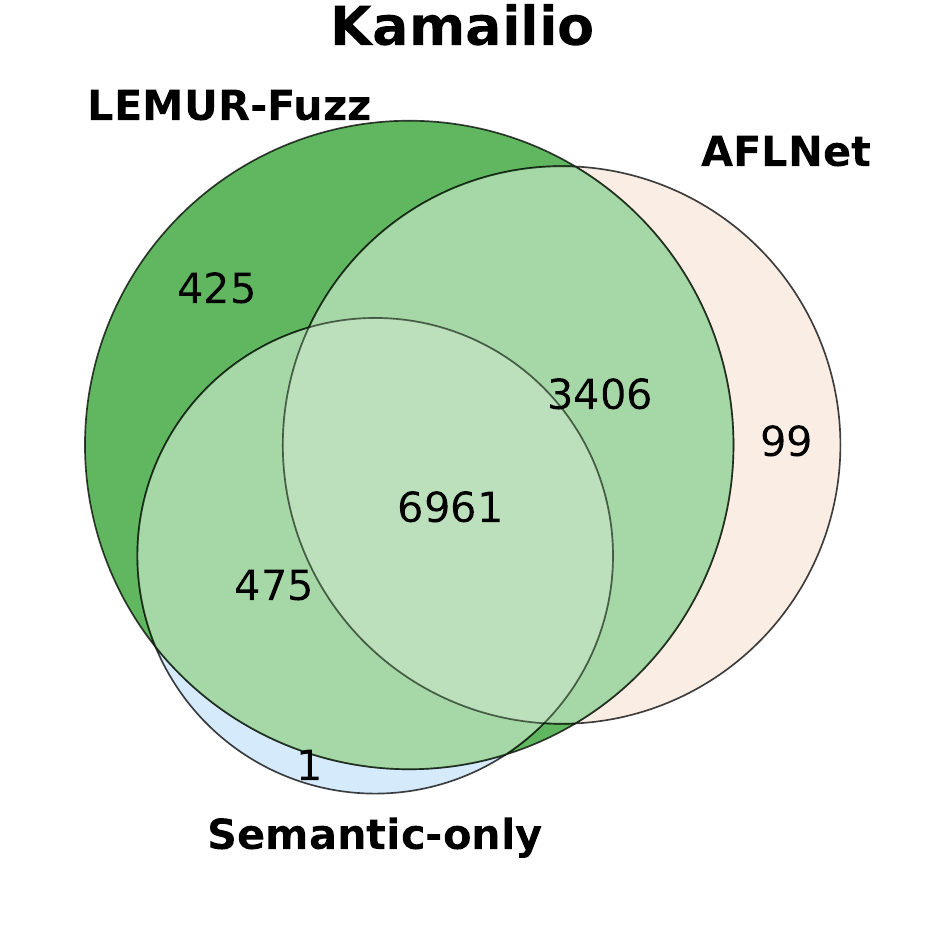}} &
    \subcaptionbox{Live555\label{fig:venn-live555}}
      {\includegraphics[width=0.24\linewidth, trim=8 8 8 35.6, clip]{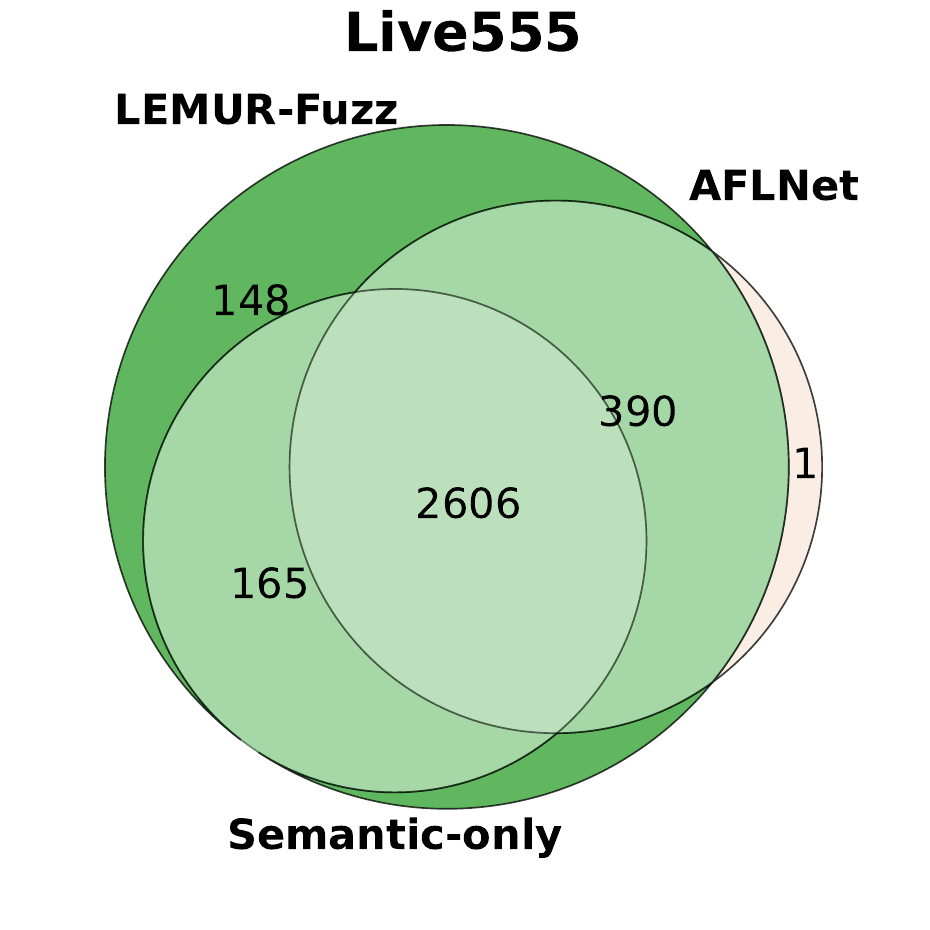}} \\

    \subcaptionbox{ProFTPD\label{fig:venn-proftpd}}
      {\includegraphics[width=0.24\linewidth, trim=8 8 8 34.4, clip]{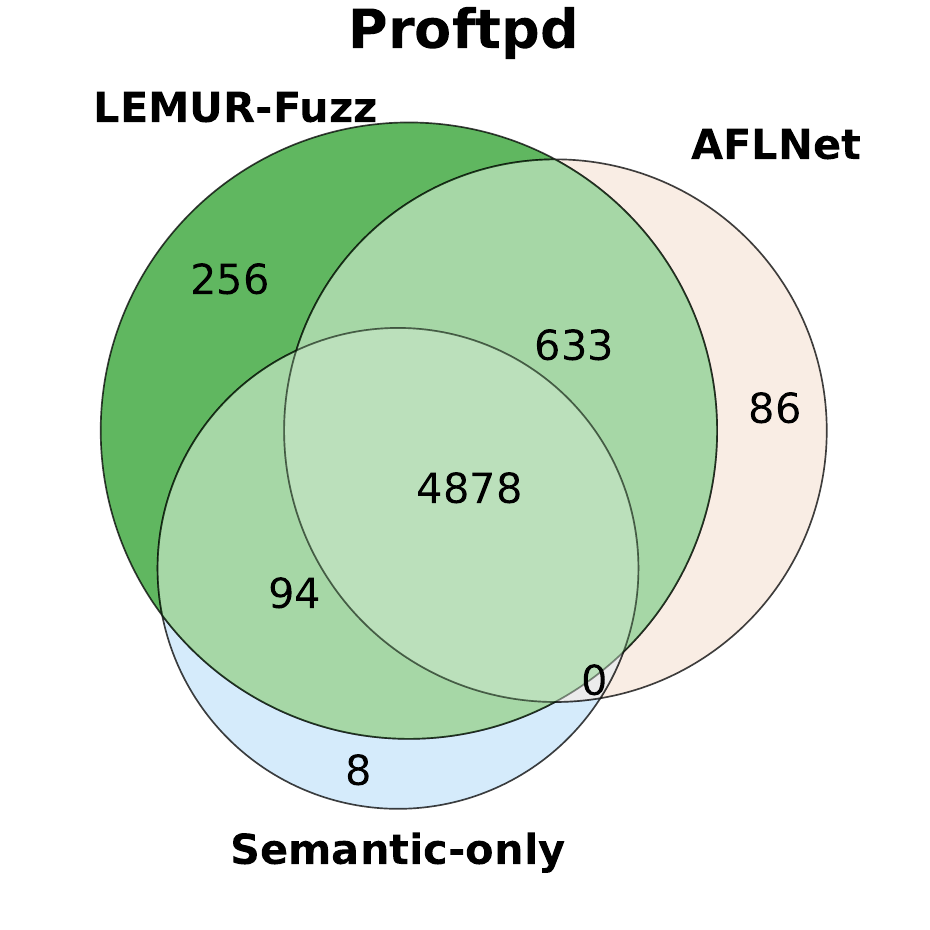}} &
    \subcaptionbox{Pure-FTPd\label{fig:venn-pureftpd}}
      {\includegraphics[width=0.24\linewidth, trim=8 8 8 36.0, clip]{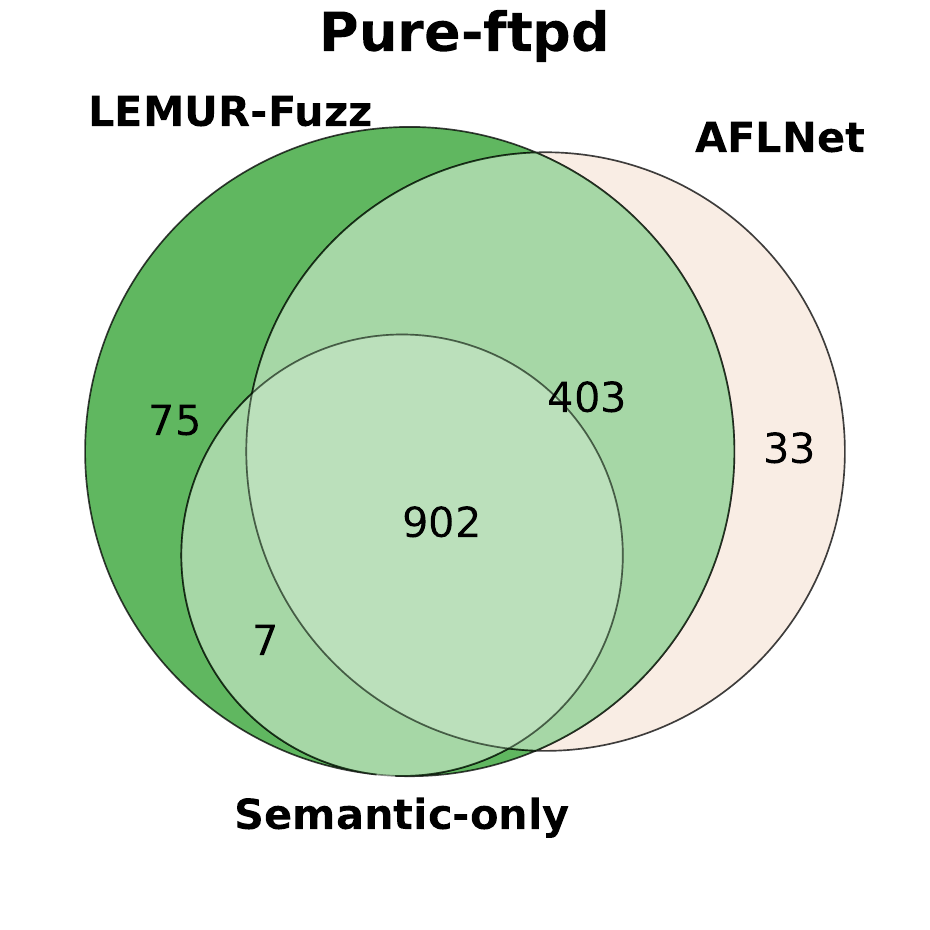}} &
    \subcaptionbox{Exim\label{fig:venn-exim}}
      {\includegraphics[width=0.24\linewidth, trim=8 8 8 30.6, clip]{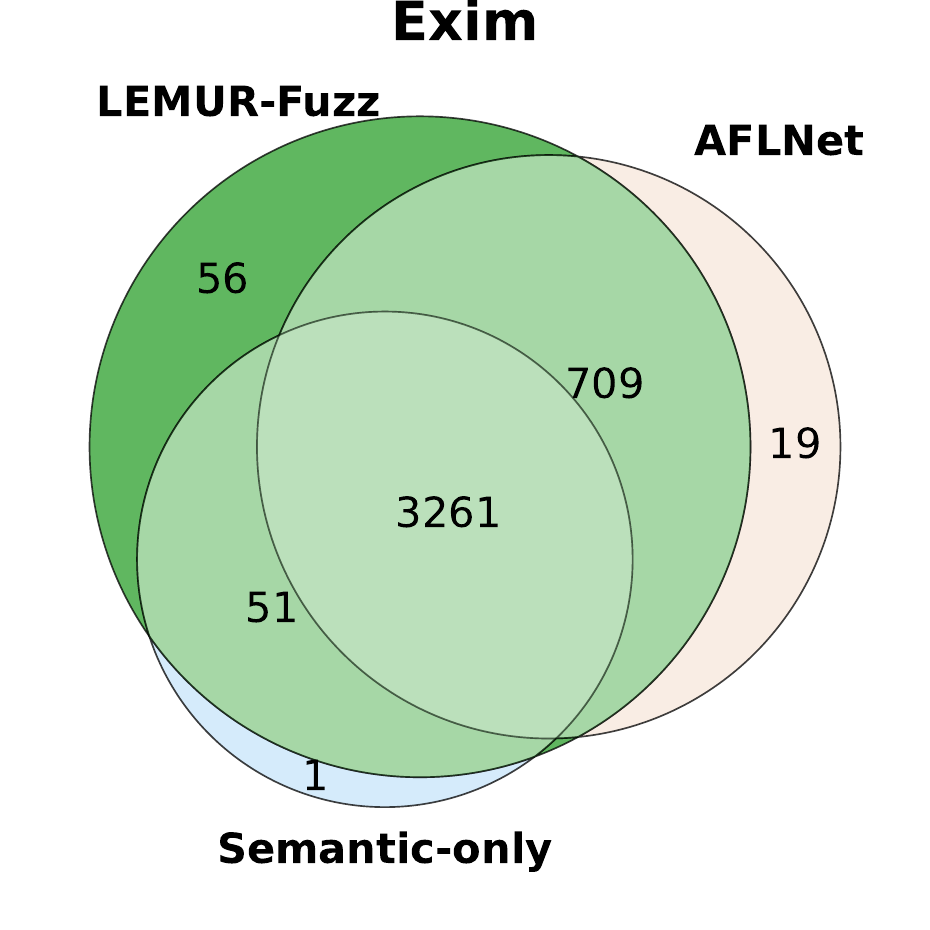}} &
    \subcaptionbox{TinyDTLS\label{fig:venn-tinydtls}}
      {\includegraphics[width=0.24\linewidth, trim=8 8 8 41.4, clip]{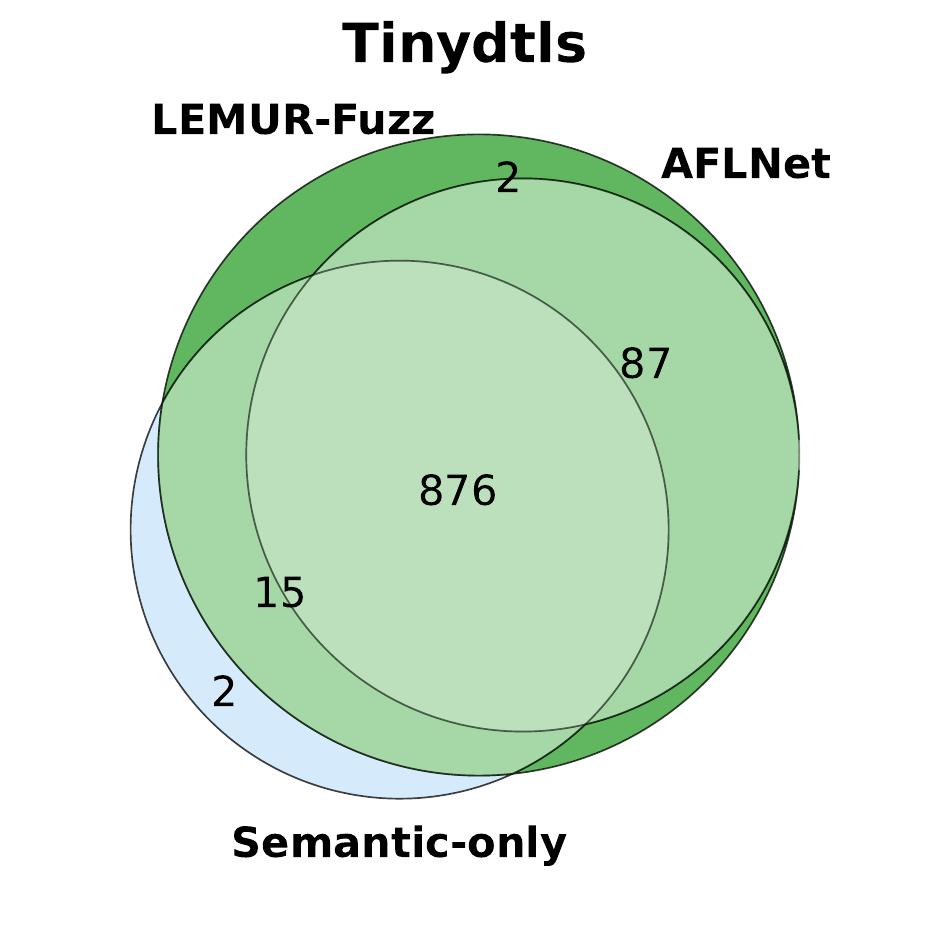}} \\

    \subcaptionbox{OpenSSL\label{fig:venn-openssl}}
      {\includegraphics[width=0.24\linewidth, trim=8 8 8 34, clip]{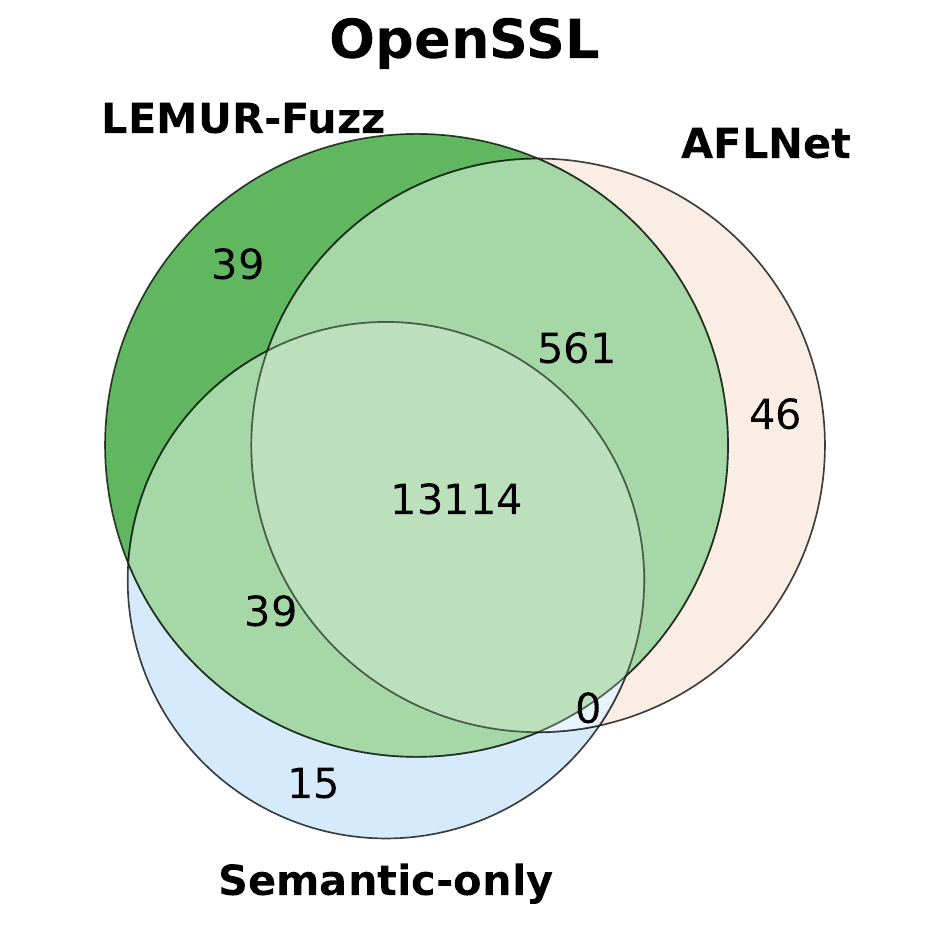}} &
    \subcaptionbox{DCMTK\label{fig:venn-dcmtk}}
      {\includegraphics[width=0.24\linewidth, trim=8 8 8 34, clip]{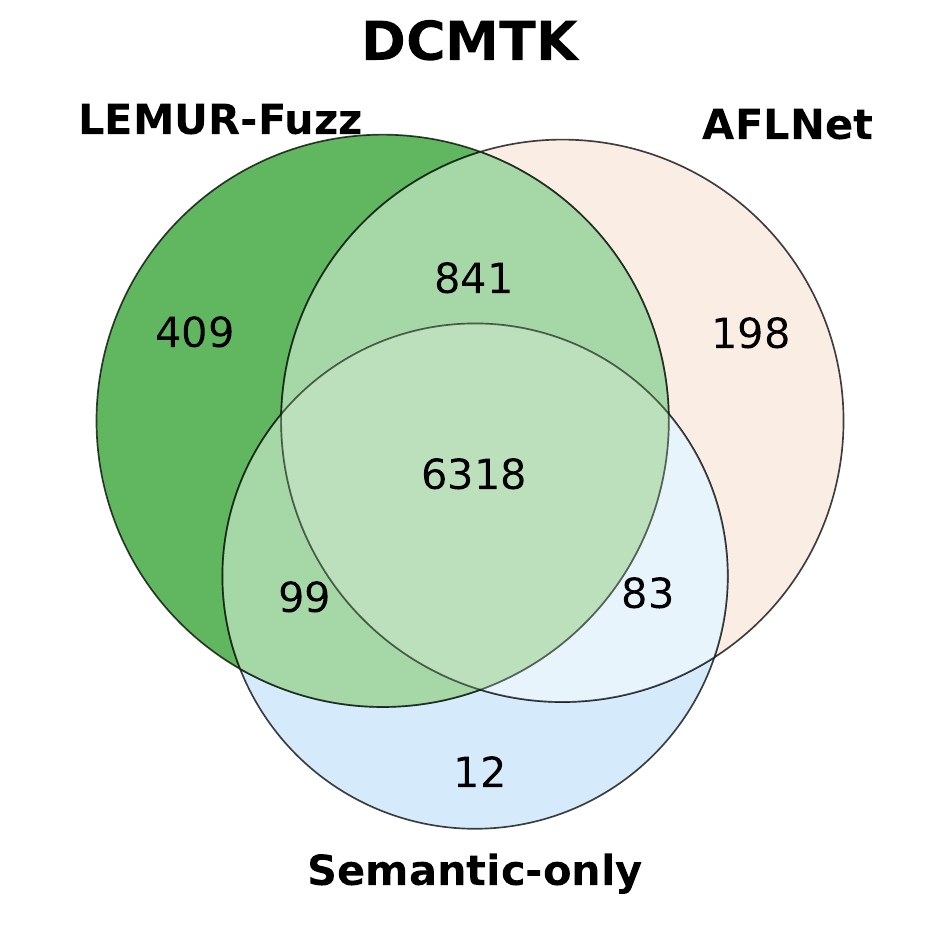}} &
    \subcaptionbox{dnsmasq\label{fig:venn-dnsmasq}}
      {\includegraphics[width=0.24\linewidth, trim=8 8 8 34, clip]{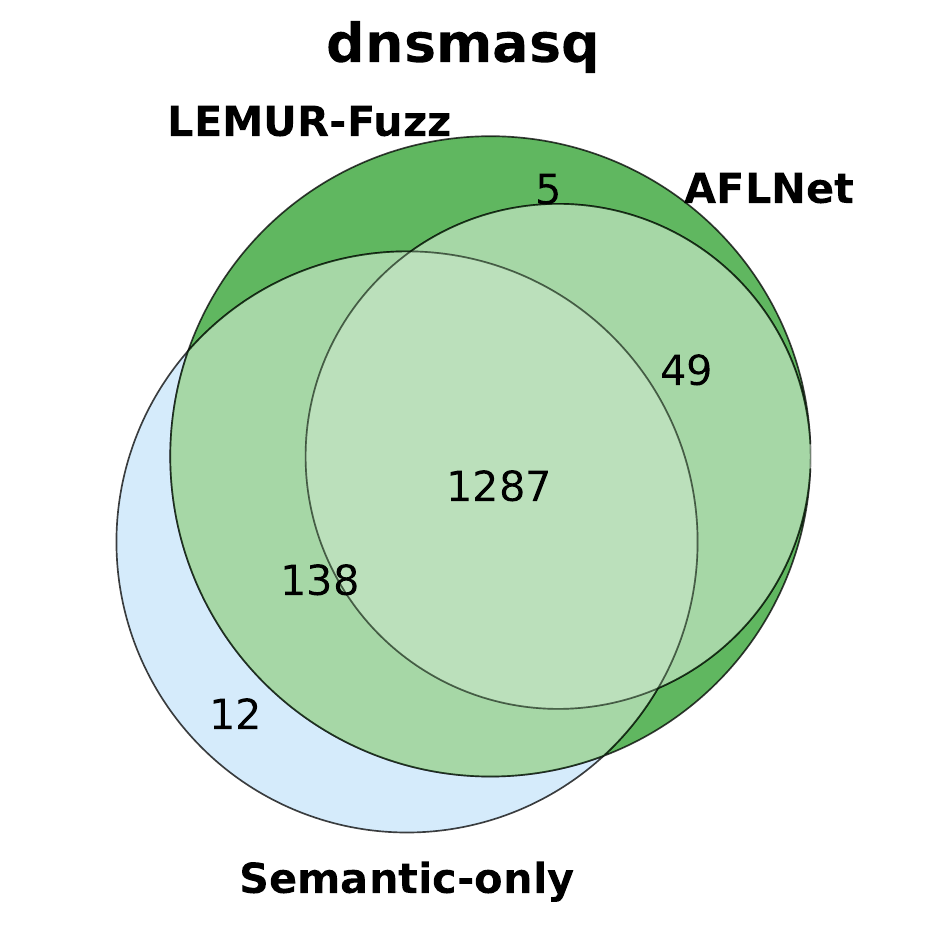}} &
    \subcaptionbox{OpenSSH\label{fig:venn-openssh}}
      {\includegraphics[width=0.24\linewidth, trim=8 8 8 8, clip]{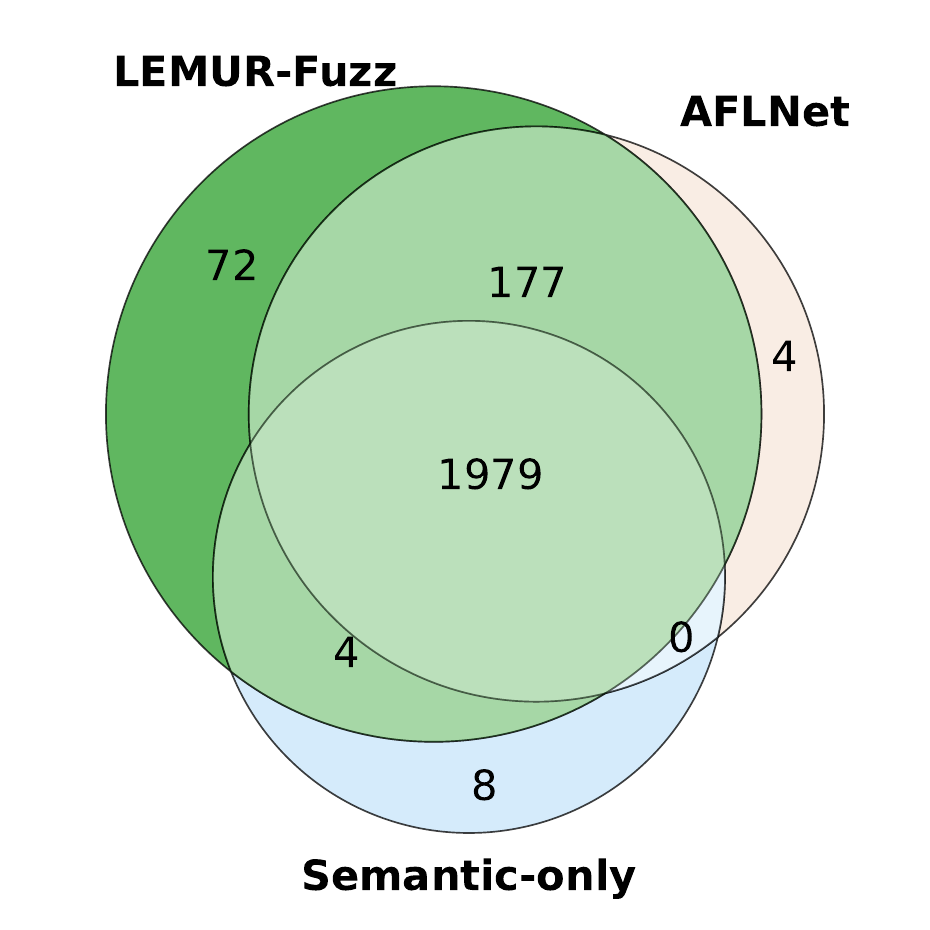}} \\
  \end{tabular}

  \caption{Venn diagrams of branches covered by LEMUR-Fuzz, AFLNet, and the Semantic-only baseline across the twelve protocol implementations. Each panel is built from the union of ten branch-level replay profiles. For OpenSSH, all three configurations use the same 2613-byte initial seed and gcov replay procedure. The panels use the same green, peach, and pale-blue style, and all regions are computed from branch identities rather than scalar averages.}

  \label{fig:venn-all}
\end{figure}

LEMUR-Fuzz augments AFLNet with a semantic-aware mutation stage.
We use \emph{Semantic-only} to denote a variant of LEMUR-Fuzz in which original AFLNet mutations are disabled and only the semantic-aware mutation stage is applied.
To justify our design choice, we compare the branches covered by LEMUR-Fuzz, Semantic-only, and AFLNet, and visualize their differences using Venn diagrams. For each implementation and configuration, we union the \emph{unique\_branch} sets from ten runs. Each Venn panel then shows the pairwise, three-way, and exclusive regions of the resulting AFLNet, LEMUR-Fuzz, and Semantic-only sets, rather than scalar coverage averages.
Beyond a large shared core, LEMUR-Fuzz covers branches missed by both AFLNet and Semantic-only across the twelve implementations within the same 24-hour budget. The LEMUR-Fuzz-only region contains 2--425 branches per implementation, averaging 152.3 branches across the benchmark. The largest regions occur for Kamailio (425 branches), DCMTK (409), and ProFTPD (256), reflecting differences in protocol state spaces and implementation structure. AFLNet and Semantic-only are also complementary across the benchmark, while the full LEMUR-Fuzz configuration combines both sources of exploration. Together, the regions in Figure~\ref{fig:venn-all} show that neither semantic-aware mutation nor havoc/splicing alone is sufficient for broad path exploration.

\subsection{RQ3: Ablation Study}
% \subsubsection{Ablation Study of the semantic-aware mutator and constraint fixer components}

LEMUR-Fuzz has two core components that directly affect the mutation stage:
(i) semantic-aware field-level mutators, and
(ii) constraint fixers.
Therefore, we design three ablation variants to isolate their individual and combined effects:
\begin{itemize}
 % \item \emph{Base}: disabling both semantic-aware mutators and constraint fixers, equivalent to AFLNet.
  \item \emph{Base}: disabling both components, equivalent to AFLNet.
 \item \emph{LEMUR-Fuzz$_{\text{Mut}}$}: enabling only semantic-aware mutators.
 \item \emph{LEMUR-Fuzz$_{\text{Fix}}$}: enabling only constraint fixers.
\end{itemize}

\begin{table}[t]
\centering
\caption{Ablation study of semantic-aware mutators and constraint fixers in LEMUR-Fuzz. Here, LEMUR-Fuzz$_{\text{Mut}}$ corresponds to the no-fixer configuration, and LEMUR-Fuzz$_{\text{Fix}}$ corresponds to no-sem.}
\label{tab:ablation_mutator_fixer}
\small
\setlength{\tabcolsep}{0pt}
\begin{threeparttable}
\begin{tabular*}{\linewidth}{@{\extracolsep{\fill}}lrrrrrrrrrrrrr@{}}
\toprule
\multirow{2}{*}{\textbf{Target}} & \multirow{2}{*}{\textbf{Base}} &
\multicolumn{4}{c}{\textbf{LEMUR-Fuzz$_{\text{Mut}}$}} &
\multicolumn{4}{c}{\textbf{LEMUR-Fuzz$_{\text{Fix}}$}} &
\multicolumn{4}{c}{\textbf{LEMUR-Fuzz}} \\
\cmidrule(lr){3-6}\cmidrule(lr){7-10}\cmidrule(l){11-14}
& & \textbf{Br.} & \textbf{Imp.} & \textbf{$p$} & \textbf{$A_{12}$ (95\% CI)}
& \textbf{Br.} & \textbf{Imp.} & \textbf{$p$} & \textbf{$A_{12}$ (95\% CI)}
& \textbf{Br.} & \textbf{Imp.} & \textbf{$p$} & \textbf{$A_{12}$ (95\% CI)} \\
\midrule
Mosquitto & 2166.2 & 2342.5 & +8.14\% & $<0.05$ & $1.00$\,[1.00,1.00] & 2248.4 & +3.79\% & $<0.05$ & $1.00$\,[1.00,1.00] & 2423.2 & +11.86\% & $<0.05$ & $1.00$\,[1.00,1.00] \\
NanoMQ & 2779.2 & 3102.7 & +11.64\% & $<0.05$ & $1.00$\,[1.00,1.00] & 2829.8 & +1.82\% & $<0.05$ & $0.82$\,[0.61,1.00] & 3141.8 & +13.05\% & $<0.05$ & $1.00$\,[1.00,1.00] \\
live555 & 2895.8 & 3090.3 & +6.72\% & $<0.05$ & $1.00$\,[1.00,1.00] & 2966.6 & +2.44\% & $<0.05$ & $0.97$\,[0.88,1.00] & 3145.0 & +8.61\% & $<0.05$ & $1.00$\,[1.00,1.00] \\
Kamailio & 9886.2 & 10140.1 & +2.57\% & $<0.05$ & $0.89$\,[0.71,0.99] & 10280.2 & +3.99\% & $<0.05$ & $0.99$\,[0.94,1.00] & 10814.8 & +9.39\% & $<0.05$ & $1.00$\,[1.00,1.00] \\
ProFTPD & 5097.8 & 5147.3 & +0.97\% & $<0.05$ & $0.79$\,[0.54,0.99] & 5180.1 & +1.61\% & $<0.05$ & $0.77$\,[0.52,0.99] & 5316.1 & +4.28\% & $<0.05$ & $0.91$\,[0.73,1.00] \\
Pure-FTPd & 1184.7 & 1227.9 & +3.65\% & $0.10$ & $0.72$\,[0.44,0.96] & 1185.7 & +0.08\% & $0.91$ & $0.48$\,[0.20,0.74] & 1275.9 & +7.70\% & $<0.05$ & $0.89$\,[0.70,1.00] \\
Exim & 3907.7 & 3923.2 & +0.40\% & $0.43$ & $0.61$\,[0.34,0.85] & 3926.6 & +0.48\% & $0.41$ & $0.61$\,[0.35,0.85] & 3938.3 & +0.78\% & $0.39$ & $0.62$\,[0.36,0.86] \\
TinyDTLS & 932.7 & 969.1 & +3.90\% & $<0.01$ & $1.00$\,[1.00,1.00] & 938.7 & +0.64\% & $0.23$ & $0.67$\,[0.41,0.89] & 958.4 & +2.76\% & $<0.05$ & $1.00$\,[1.00,1.00] \\
dnsmasq & 1335.5 & 1457.1 & +9.11\% & $<0.05$ & $1.00$\,[1.00,1.00] & 1433.0 & +7.30\% & $<0.05$ & $1.00$\,[1.00,1.00] & 1479.7 & +10.80\% & $<0.05$ & $1.00$\,[1.00,1.00] \\
DCMTK & 6495.5 & 6862.7 & +5.65\% & $<0.05$ & $0.87$\,[0.68,1.00] & 6525.2 & +0.46\% & $0.20$ & $0.47$\,[0.18,0.78] & 7173.5 & +10.44\% & $<0.05$ & $0.98$\,[0.91,1.00] \\
OpenSSL & 13469.1 & 13531.8 & +0.47\% & $<0.05$ & $0.80$\,[0.55,1.00] & 13514.1 & +0.33\% & $<0.05$ & $0.68$\,[0.40,0.92] & 13547.0 & +0.58\% & $<0.05$ & $0.78$\,[0.54,0.97] \\
OpenSSH & 2131.6 & 2156.5 & +1.17\% & $<0.05$ & $0.91$\,[0.75,1.00] & 2139.1 & +0.35\% & $0.17$ & $0.69$\,[0.43,0.91] & 2193.4 & +2.90\% & $<0.05$ & $1.00$\,[1.00,1.00] \\
\midrule
\textbf{Average} & 4356.8 & 4495.9 & \textbf{+4.53\%} & -- & - & 4430.6 & \textbf{+1.94\%} & -- & - & 4617.3 & \textbf{+6.93\%} & -- & - \\
\bottomrule
\end{tabular*}
\begin{tablenotes}[flushleft]
\footnotesize
\item[] Effect sizes are reported using Vargha--Delaney $A_{12}$ with 95\% bootstrap confidence intervals (CI), rounded to two decimals. All rows use the preserved run-level samples for the corresponding implementation and configuration. Aggregate $p$-values and AVG-row $A_{12}$ values are not reported. Br. and Imp. denote branch coverage and relative improvement, respectively.
\end{tablenotes}
\end{threeparttable}
\end{table}

The ablation results cover semantic-aware mutators and constraint fixers across the twelve implementations. Compared with the \textit{Base} (AFLNet), enabling semantic-aware mutation alone improves branch coverage on all targets, with gains from 0.40\% to 11.64\% (4.53\% on average). These results confirm that field-level semantic-aware mutation improves input validity over byte- and region-level mutation.
The fixer-only variant also improves coverage, but with smaller gains: LEMUR-Fuzz$_{\text{Fix}}$ achieves a 1.94\% average improvement, with target-level gains from 0.08\% to 7.30\%. This suggests that constraint fixing can help even without semantic-aware mutation, but its standalone effect is limited because byte-level mutations often corrupt message structures before inputs can be parsed and repaired. A fixer enforces relationships among fields, but it does not by itself generate diverse, semantically meaningful field values. Semantic-aware mutation therefore provides the rich protocol values needed to reach new states, while the fixer restores dependent-field consistency so that these inputs remain executable and can explore deeper code.

\pagebreak
As Table~\ref{tab:ablation_mutator_fixer} shows, the full configuration (LEMUR-Fuzz) performs best on average, improving coverage by 6.93\%, compared with 4.53\% from mutation alone and 1.94\% from fixing alone. This shows that the two components are complementary: mutators generate protocol-valid inputs, while fixers repair cross-field inconsistencies that help inputs reach deeper paths.

\subsection{RQ4: Message Validity}
\label{sec:rq4_validity}

\begin{table}[t]
\centering
\normalsize
\caption{Grammar, semantic, and execution success rates at 24 hours (10 runs averaged).}
\label{tab:validity-results}
\setlength{\tabcolsep}{2.0pt}

\begin{threeparttable}
\begin{tabular*}{\linewidth}{@{\extracolsep{\fill}}lrrrrrrrrr@{}}
\toprule
\multirow{2}{*}{\textbf{Target}} &
\multicolumn{3}{c}{\textbf{AFLNet}} &
\multicolumn{3}{c}{\textbf{Semantic-aware mut.}} &
\multicolumn{3}{c}{\textbf{Inc. over AFLNet}} \\
\cmidrule(lr){2-4}\cmidrule(lr){5-7}\cmidrule(l){8-10}
\multicolumn{1}{l}{} &
\textbf{gram} & \textbf{sem} & \multicolumn{1}{c}{\textbf{exec}} &
\textbf{gram} & \textbf{sem} & \multicolumn{1}{c}{\textbf{exec}} &
\textbf{gram} & \textbf{sem} & \textbf{exec} \\
\midrule
Mosquitto
& 12.4\% & 10.8\% & 3.9\%
& 89.7\% & 64.3\% & 38.0\%
& 623.4\% & 495.4\% & 874.4\% \\

NanoMQ
& 46.2\% & 45.5\% & 7.5\%
& 96.6\% & 95.9\% & 83.4\%
& 109.1\% & 110.8\% & 1012.0\% \\

Kamailio
& 70.0\% & 45.9\% & 45.9\%
& 95.1\% & 77.0\% & 68.8\%
& 35.9\% & 67.8\% & 49.9\% \\

live555
& 62.3\% & 47.0\% & 11.8\%
& 80.0\% & 72.7\% & 34.9\%
& 28.4\% & 54.7\% & 195.8\% \\

ProFTPD
& 97.9\% & 45.2\% & 33.5\%
& 98.3\% & 71.0\% & 53.2\%
& 0.4\% & 57.1\% & 58.8\% \\

Pure-FTPd
& 84.0\% & 28.8\% & 25.0\%
& 91.5\% & 62.3\% & 57.1\%
& 8.9\% & 116.3\% & 128.4\% \\

Exim
& 44.2\% & 28.2\% & 25.0\%
& 77.2\% & 55.4\% & 52.2\%
& 74.7\% & 96.5\% & 108.8\% \\

TinyDTLS
& 69.6\% & 27.2\% & 19.6\%
& 80.5\% & 45.6\% & 33.9\%
& 15.7\% & 67.6\% & 73.0\% \\

OpenSSL
& 84.7\% & 19.4\% & 2.1\%
& 96.4\% & 79.9\% & 20.2\%
& 13.8\% & 311.9\% & 861.9\% \\

DCMTK
& 64.7\% & 43.1\% & 3.9\%
& 83.6\% & 68.1\% & 31.5\%
& 29.2\% & 58.0\% & 707.7\% \\

dnsmasq
& 30.8\% & 8.7\% & 1.2\%
& 98.1\% & 69.9\% & 32.4\%
& 218.5\% & 703.4\% & 2600.0\% \\

OpenSSH
& 85.7\% & 64.3\% & 5.8\%
& 97.2\% & 80.0\% & 20.0\%
& 13.4\% & 24.4\% & 244.8\% \\
\midrule
\textbf{AVG} & \textbf{62.7\%} & \textbf{34.5\%} & \textbf{15.4\%} & \textbf{90.4\%} & \textbf{70.2\%} & \textbf{43.8\%} & \textbf{+97.6\%} & \textbf{+180.3\%} & \textbf{+576.3\%} \\
\bottomrule
\end{tabular*}

\begin{tablenotes}[flushleft]
\footnotesize
\item[] The last three columns report relative increases over AFLNet. AVG is the unweighted arithmetic mean of the twelve displayed values in each column; the AVG increases are means of per-target increases, not increases computed from the mean success rates.
\end{tablenotes}
\end{threeparttable}

\end{table}

RQ4 evaluates whether the semantic-aware mutation stage of LEMUR-Fuzz improves the validity of mutated protocol messages during fuzzing.
We instrumented twelve protocol implementations to measure how far each mutated $M_2$ message proceeds in the server-side pipeline. We use three common stages: \emph{grammar success} (\emph{gram}) means format-level parsing succeeds; \emph{semantic success} (\emph{sem}) means further field-level checks such as required fields, value ranges, and cross-field consistency succeed; and \emph{execution success} (\emph{exec}) means the message is accepted and processed. 
The instrumentation was manually performed in the corresponding server-side handler functions.
For example, in live555, \emph{gram} is recorded after \code{parseRTSPRequestString()} parses the request, \emph{sem} after command-specific checks succeed (e.g., locating the media session for DESCRIBE or validating the track and \code{Transport} header for SETUP), and \emph{exec} when a successful RTSP response such as \code{setRTSPResponse("200 OK", ...)} is generated.

As shown in Table~\ref{tab:validity-results}, LEMUR-Fuzz's semantic-aware mutation stage improves all three ratios across all twelve implementations. Over the 24 hours, average grammar success increases from 62.7\% to 90.4\%, semantic success from 34.5\% to 70.2\%, and execution success from 15.4\% to 43.8\%. The gains are especially large on implementations that reject many malformed messages: Mosquitto improves from 12.4\% to 89.7\% in grammar success and from 3.9\% to 38.0\% in execution success, while NanoMQ improves from 7.5\% to 83.4\% in execution success. These results show that semantic-aware mutation better preserves message validity and field constraints, helping mutated messages explore deep protocol states.

\section{Bug Case Study}

\begin{figure}[t]
    \centering
    \includegraphics[width=\linewidth]{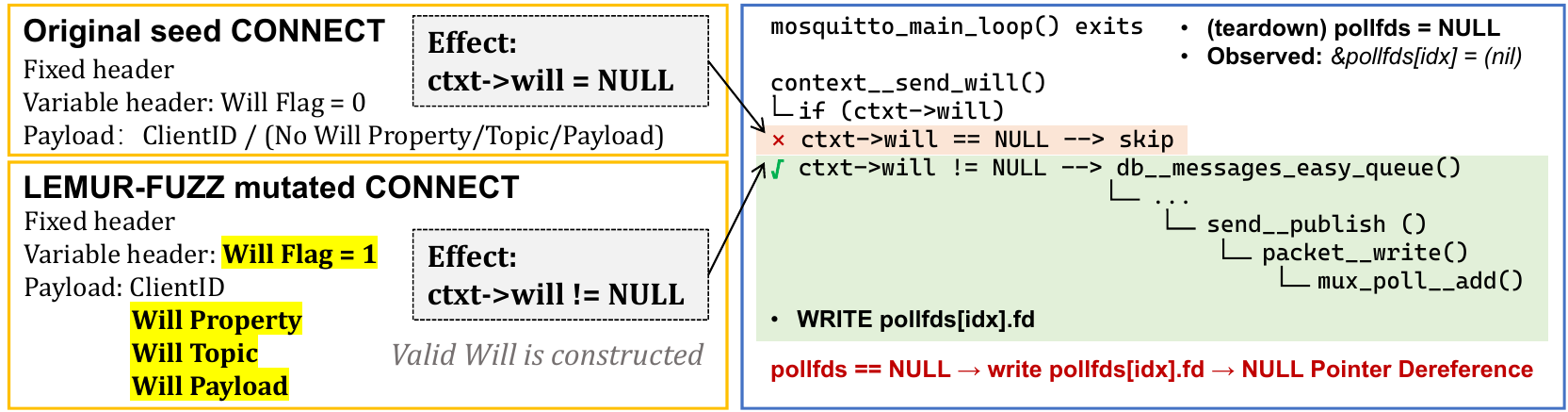}
    \caption{The semantic-aware CONNECT mutation of LEMUR-Fuzz constructs a valid Will-enabled MQTT session to trigger a teardown-time NULL dereference in Mosquitto.}
    % \vspace{-0.5em}
    \label{fig:mosquitto-bug}
\end{figure}

% \subsection{Case Study 1: Null-Dereference in Mosquitto}
% We use Mosquitto to illustrate how LEMUR-Fuzz satisfy strict semantic validations, drive the broker into deeper stateful logic and ultimately trigger subtle bugs that are rarely reachable with conventional fuzzing.
Figure~\ref{fig:mosquitto-bug} illustrates how LEMUR-Fuzz triggers a previously unknown crash in Mosquitto by enabling an execution path that is unreachable for other conventional fuzzers such as AFLNet.
This bug is confirmed and fixed by Mosquitto developers after we reported it.

\noindent\textbf{Bug Triggering Mechanism.}
The crash occurs during broker termination, when Mosquitto attempts to deliver a client \emph{Will} message.
If a client configures a Will during the CONNECT handshake, the broker invokes
\func{context\_send\_will()} to enqueue the Will as a normal PUBLISH.
This delivery path eventually reaches the functions
\func{packet\_write()} and \func{mux\_poll\_add()}, which update the poll-based I/O state by modifying entries in the \var{pollfds} array (e.g., writing to \code{pollfds[idx].\allowbreak fd} to register file descriptors for I/O events).
However, at termination time, the broker’s polling structures have already been torn down, and the pointer
\var{pollfds} becomes unavailable.
Specifically, after control exits \func{mosquitto\_main\_loop()}, the \var{pollfds} array is freed.
As a result, \func{mux\_poll\_add()} attempts to write to \code{pollfds[idx].\allowbreak fd} while
\var{pollfds} is \code{NULL}, leading to a NULL-pointer dereference and a crash, as confirmed by AddressSanitizer.

\noindent\textbf{Why Only LEMUR-Fuzz Finds The Bug.}
Reaching this crash requires a valid Will configuration. If a CONNECT message does not include Will fields, \code{ctxt->will == NULL} holds and the Will-delivery path is skipped. Using the same seed, AFLNet fails to trigger this bug in 10 runs of 24-hour fuzzing because general-purpose mutations often violate dependencies between the Will flag and Will-related payload fields. In contrast, LEMUR-Fuzz generates a protocol-compliant CONNECT message with a valid Will configuration, initializes \code{ctxt->will}, passes the semantic gate, and reaches the teardown-time path where the crash is triggered.

\section{Discussion}
\label{sec:discussion}

\noindent\textbf{Manual Effort.}
% LEMUR-Fuzz currently adopts a \emph{semi-automated} workflow while manual inspection remains necessary at certain stages. In particular, manually checking the correctness of protocol message templates is critical, as inaccuracies at this stage may propagate to downstream components.
% We introduce a set of automated correctness-checking mechanisms to assess the usability of LLM-generated code. These tests effectively identify many implementation errors. When failures occur, human intervention is still required to inspect and repair the generated artifacts. In practice, adding support for a new protocol typically requires one to two days of effort, depending on protocol complexity and the quality of the underlying language model.
% Overall, LEMUR-Fuzz significantly reduces the manual effort compared to fully handcrafted protocol fuzzers, while retaining human oversight at critical correctness-checking points to ensure correctness and reliability.
In our approach, preparing the four LLM-generated components for fuzzing a protocol may require some manual work. As the foundation of the other components, LLM-generated message templates need manual checking, which takes roughly 10 minutes per protocol on average. 
For the errors revealed by quality checkers in LLM-generated parsers, reassemblers, and fixers,
we perform manual repair.
The error-revealing test cases help localize the error. 
In our experience, repairing these components for a protocol takes 30 minutes to 2 hours (depending on protocol complexity and components' code sizes). 
For mutators, we do not manually repair those that fail the checks; we discard them instead.
Note that the preceding cost of supporting a protocol is a one-time cost.

\noindent\textbf{Effectiveness of Quality Checkers.}
 \label{Effectiveness of Quality Checkers}
The quality checkers in our approach cannot guarantee the correctness of the LLM-generated components, but provide empirically effective quality checking.
Following the definition in Section~\ref{sec:rq4_validity}, we use message \emph{execution success rate} as the metric to assess the effectiveness of quality checkers. A message is successfully executed if it is accepted and processed by the protocol server. Intuitively, given a successfully executed message, a low-quality parser/reassembler may corrupt it during parsing and reassembling, causing the resultant message to fail execution; likewise, a low-quality mutator may generate corrupted messages that fail execution. From AFLNet's seed pools over 24-hour fuzzing, we collected all seeds that can be successfully executed by the corresponding protocol servers, obtaining 89,202 seeds from a separate validation subset. We use these seeds to assess the checkers.

With the metamorphic checker, 
%For parser/reassembler checking, we compare the execution success rate of reassembled messages produced by LLM-generated parsers/reassemblers before checking with those produced by parsers/reassemblers retained after checking. 
the average execution success rate of parser-reassembler pairs improves from 46.40\% to 94.11\%, with gains on all targets. The largest improvements are 95.54\% on Kamailio, 59.33\% on Mosquitto, and 56.06\% on live555.
With the parseability checker, 
% For mutator checking, we compare the execution success rate of mutated messages produced by LLM-generated mutators before checking with those produced by mutators retained after checking. 
% We run 100 mutation trials for each mutator to account for randomness. 
the average execution success rate of mutators improves from 71.36\% to 89.72\%. The largest improvements are 45.12\% on Mosquitto, 41.69\% on Exim, 33.27\% on ProFTPD, and 19.52\% on live555.
We did not quantify the effectiveness of the constraint-violation checker using execution success rate, because a repaired message may still be rejected for reasons unrelated to the cross-field constraint. Thus, server execution success is not an appropriate standalone measure for this checker.

\noindent\textbf{Novelty of LEMUR-Fuzz's Semantic-Aware Mutators.}
LEMUR-Fuzz's semantic-aware mutators are generated based on protocol-specific field semantics specified in RFCs. 
Such mutators have not been considered by prior protocol fuzzers.
For example, mutation-based fuzzers like AFLNet only use general-purpose mutators. Generation-based fuzzers like Peach~\cite{peach} and Boofuzz~\cite{boofuzz} have \emph{type-aware} but no \emph{semantic-aware} mutators.
Specifically, Peach could do mutation based on some specific type (e.g., integer, string, or byte-array) of a field: generating boundary values for integer fields, changing the length or content of string fields, and altering raw bytes in byte-array fields. 
Such type-aware mutators cannot achieve the effect of LEMUR-Fuzz's semantic-aware mutators. For example, LEMUR-Fuzz could generate meaningful values of identifiers and topics specific to MQTT (see examples in Section~\ref{sec:semantic-aware-mutators}). 
Indeed, such semantic-aware mutators could also enhance other fuzzers. 
In a separate Peach study, adding LEMUR-Fuzz's semantic-aware mutators and constraint fixers improved branch coverage by 5.66\% on average, with the largest gains on Mosquitto (11.16\%) and Pure-FTPd (9.48\%).

\noindent\textbf{Impact on Testing Efficiency.}
Although our pipeline adds parse--mutate--fix--reassemble steps, the throughput impact is negligible. AFLNet spends 31\,\textmu s per execution on mutation, while our approach spends 56\,\textmu s; in contrast, server interaction takes about 140\,ms per execution, which dominates end-to-end latency and makes the added local overhead insignificant.
% Despite introducing parsing, repair, and reassembly, our approach does not impose a meaningful throughput penalty in practice. We break down the per-execution latency into (i) local time spent generating the next input and (ii) time spent interacting with the server (sending inputs, receiving responses, and waiting on network/timeouts).
% In our measurements, AFLNet spends 31\,\textmu s per execution on mutation on average, while our approach spends 56\,\textmu s on average due to the added parse--mutate--fix--reassemble steps. By contrast, server interaction takes 140\,ms per execution on average, dwarfing the 31--56\,\textmu s spent on mutation. Therefore, the overhead is negligible in the end-to-end fuzzing loop and does not affect overall throughput.

\section{Related Work}

\noindent\textbf{Protocol Fuzzing.}
Existing protocol fuzzers fall into two major categories: mutation-based and generation-based.
Mutation-based protocol fuzzers~\cite{aflnet,natella2022stateafl,snapfuzz,nsfuzz,aflnwe,aflnet++,DYFuzzing,Frankenstein} generate new inputs by mutating initial seeds with general-purpose mutators. 
% Among them, the two most prominent architectural lineages are AFL-based and libFuzzer-based protocol fuzzers. AFL-based protocol fuzzers \cite{aflnet,natella2022stateafl,snapfuzz,nsfuzz,aflnwe,aflnet++} inherit AFL’s forkserver-driven, process-per-test execution model and bitmap-based coverage feedback.
AFLNet~\cite{aflnet} is a pioneer mutation-based fuzzer which introduces stateful coverage-guided protocol fuzzing. StateAFL~\cite{natella2022stateafl} further improved AFLNet with state coverage feedback.
%  NSFuzz enhances protocol fuzzing by adopting a more precise variable-based state representation and an efficient interaction synchronization mechanism~\cite{nsfuzz}. 
%  In parallel, libFuzzer-based protocol fuzzers inherit libFuzzer's in-process execution model, SanitizerCoverage instrumentation hooks, and built-in corpus management. SGFuzz~\cite{sgfuzz} extends libFuzzer to stateful protocols by automatically discovering programmatic protocol states (e.g., enum-typed state variables) and using their runtime transitions as state-coverage feedback to guide mutation.
%Despite these advances in state awareness, most mutation-based protocol fuzzers still rely on the
However, their byte-level, general-purpose mutators would likely generate invalid protocol messages.
% In contrast, \TechniqueName\ augments protocol fuzzers with semantic-aware mutators and constraint fixers, enabling the generation of more valid protocol messages.
Generation-based protocol fuzzers~\cite{peach,boofuzz,greyhound,kif,snooze,zfuzzer,atlidakis2019restler} generate inputs by leveraging manually constructed protocol data/state models.
%(e.g., Peach~\cite{peach}, Boofuzz~\cite{boofuzz}) 
% Other fuzzers such as Zfuzzer~\cite{zfuzzer} and RESTler~\cite{atlidakis2019restler} also follow this fuzzing paradigm.
% Zfuzzer augments the black box generation approach of Boofuzz with code coverage guidance, yet it still relies on manually modeling the Zigbee protocol to generate well-formed messages~\cite{zfuzzer}. In a different direction, RESTler generates grammar conformant request sequences with producer–consumer dependencies from API schemas, effectively extending grammar awareness from individual messages to stateful API workflows~\cite{atlidakis2019restler}. While this reduces the need for manual modeling of syntax and dependencies, its generalizability is limited, as it primarily targets RESTful APIs rather than general network protocols. Overall, the effectiveness of generation-based protocol fuzzers often hinges on the availability and accuracy of protocol-specific data/state models: constructing and maintaining these models incurs substantial manual effort and requires significant protocol expertise. Meanwhile, approaches that leverage higher level specifications can alleviate manual modeling, but they are frequently tailored to specific targets or interface paradigms, which constrains their applicability across diverse protocols. 
% In contrast, \TechniqueName\ leverages LLMs to automatically extract protocol syntax and semantic constraints from natural-language specifications and generate semantic-aware mutators, reducing manual modeling overhead while providing protocol-specific mutation capabilities.
Peach's built-in mutators are mainly type-aware; encoding protocol-specific field semantics requires additional user-defined modeling or logic. LEMUR-Fuzz fills this gap by generating semantic-aware mutators for deeper exploration while retaining general-purpose mutation for edge cases.

\noindent\textbf{LLM-assisted Protocol Fuzzing.}
% A growing body of work has explored leveraging LLMs to enhance fuzzing, either by generating inputs or diversifying seed corpora, or by using LLMs directly as the mutation mechanism to rewrite or generate inputs end-to-end~\cite{fuzz4all,fuzzgpt,whitefox,chatafl,psg,codex,titanfuzz,chatfuzz}.
% However, these techniques cannot be directly applied to protocol fuzzing, as they overlook the inter-field and cross-field constraints unique to protocol messages.
In the protocol fuzzing domain, several recent works have applied LLM to improve protocol fuzzing~\cite{chatafl,ma2024one,cheng2024msfuzz,zhang2025statepre,semfuzz,zfuzzer}. 
ChatAFL~\cite{chatafl} integrates an LLM with AFLNet to extract protocol grammars and generate message sequence to bootstrap the initial seed
corpus and escape coverage plateaus.
LLMIF~\cite{llmif} augments an LLM with the Zigbee protocol specification to perform two tasks, i.e., protocol information extraction and device response reasoning, and generates spec-conformant Zigbee messages and interprets device replies.
BSFuzzer~\cite{bsfuzzer} uses LLMs to extract state machines and packet
semantics from the Bluetooth Core Specification, and generate message sequences for detecting protocol logic flaws.
SemFuzz~\cite{semfuzz} uses LLMs to extract semantic rules from RFCs and generates rule-violating test cases to expose semantic bugs. Unlike these works, LEMUR-Fuzz uses LLMs to generate protocol-specific semantic-aware mutators, rather than input grammars or message sequences. Semantic-Over-Syntax~\cite{semantics-over-syntax} likewise targets semantic-violation testing by generating syntactically valid but semantically inconsistent 5G RRC messages; its goal differs from LEMUR-Fuzz, which uses semantic-aware mutation and constraint fixing to preserve executable protocol messages while exploring new states.

% Beyond the protocol domain, existing semantic-aware mutation approaches mainly use semantic knowledge to generate test inputs\cite{CLFuzz,DBLP:conf/icse/Hatfield-DoddsD22,DBLP:journals/corr/abs-2511-03995}, or use semantic context to guide mutation scheduling\cite{DBLP:conf/ccs/PanL0JJWYWW21}.
% In contrast, LEMUR-Fuzz focuses on transforming protocol semantics into executable mutation operators, enabling semantic information to directly define how inputs are systematically modified.
Beyond the protocol domain, semantics-aware mutation has been explored in several representative lines of work.
(i) some fuzzers incorporate semantic knowledge during input generation, using generators, schemas, or LLM prompts to describe and produce valid inputs~\cite{DBLP:conf/issta/PadhyeLSPT19,DBLP:conf/icse/Hatfield-DoddsD22,CLFuzz,DBLP:journals/corr/abs-2511-03995}.
(ii) some incorporate semantic knowledge into scheduling, guiding the fuzzer to select structurally consistent fragments for subsequent mutation~\cite{DBLP:conf/ccs/PanL0JJWYWW21}.
(iii) a more traditional line incorporates semantic knowledge into mutation operators, defining grammar-aware transformations over parse trees, grammar automata, or learned grammars to preserve syntactic validity~\cite{superion,nautilus,gramatron,pcsg-fuzzer}.
% LEMUR-Fuzz goes one level above formal grammar and turns protocol-level semantic rules, such as the semantic meaning of each field (e.g., a topic, a URL, or an identifier) and cross-field dependencies, into executable mutation operators, so that each mutation is not only well-formed but also directly targets a specific semantic property of the protocol.
These approaches are effective at maintaining well-formed inputs, but they do not directly address protocol-level semantics, where validity often depends on the meaning of individual fields (e.g., a topic, a URL, or an identifier) and dependencies across fields. LEMUR-Fuzz targets this gap by turning protocol-level semantic rules into executable mutation operators, making mutated inputs more likely to be both well-formed and semantically valid.
MetaMut~\cite{MetaMut} leverages LLMs to generate specialized mutators for compiler fuzzing. However, it is designed around compiler-input grammars and cannot be directly applied to protocol fuzzing or model protocol-level semantics. Its generated mutators are checked automatically for compilation and execution, while their functional correctness still requires manual inspection; LEMUR-Fuzz instead validates its generated components through an automated quality-checking pipeline with three checkers.

\newcommand{\supported}{\tikz[baseline=-0.35ex]{\fill (0,0) circle (0.55ex);}}
\newcommand{\partialsupported}{\tikz[baseline=-0.35ex]{\fill (0,0.55ex) arc (90:270:0.55ex) -- cycle; \draw (0,0) circle (0.55ex);}}
\newcommand{\notsupported}{\scalebox{1.5}{$\times$}}

\begin{table*}[t]
\centering
\caption{Comparison between LEMUR-Fuzz, Semantic-Over-Syntax, Peach, ChatAFL, and MetaMut.}
\label{tab:related-positioning}
\small
\setlength{\tabcolsep}{2pt}
\begin{threeparttable}
\begin{tabularx}{\dimexpr\textwidth-8pt\relax}{@{}>{\bfseries\raggedright\arraybackslash}p{2.25cm} >{\raggedright\arraybackslash}p{3.2cm} *{4}{>{\raggedright\arraybackslash}X}@{}}
\toprule
\textbf{Dimension} &
\makecell[c]{\textbf{LEMUR-Fuzz}} &
\makecell[c]{\textbf{Semantic-Over-}\\\textbf{Syntax}~\cite{semantics-over-syntax}} &
\makecell[c]{\textbf{Peach}~\cite{peach}} &
\makecell[c]{\textbf{ChatAFL}~\cite{chatafl}} &
\makecell[c]{\textbf{MetaMut}~\cite{MetaMut}} \\
\midrule
Key idea &
Semantic-aware mutation for deep states; AFLNet-style mutation for edge cases. &
Syntax-valid semantic violations. &
Model-guided, type-aware mutation. &
LLM-generated grammars and messages. &
LLM-generated compiler mutators. \\
\arrayrulecolor{gray!45}\cmidrule(lr){1-6}\arrayrulecolor{black}
Intra-message constraint handling &
\supported\ Semantic mutators + fixers &
\supported\ Violation generation &
\partialsupported\ Custom data models &
\partialsupported\ LLM-guided messages &
- \\
\arrayrulecolor{gray!45}\cmidrule(lr){1-6}\arrayrulecolor{black}
Inter-message constraint handling &
\supported\ Fixers &
\notsupported &
\partialsupported\ Custom state models &
\partialsupported\ LLM-guided sequences &
- \\
\arrayrulecolor{gray!45}\cmidrule(lr){1-6}\arrayrulecolor{black}
Validation of LLM outputs &
Metamorphic, parseability, and fixer tests &
Schema checks &
- &
Run multiple times &
Compile/run checks and manual inspection \\
\arrayrulecolor{gray!45}\cmidrule(lr){1-6}\arrayrulecolor{black}
Inputs for LLMs &
Protocol specification &
Protocol specification + ASN.1 schema &
- &
Protocol name + seeds &
AST + mutator templates \\
\bottomrule
\end{tabularx}
\begin{tablenotes}[flushleft]
\footnotesize
\item[] "\supported\ ", "\partialsupported\ ", "\notsupported\ " denote that the work \underline{can}, \underline{can only partially}, or \underline{cannot} handle the dimensions and "-" is not applicable.
\end{tablenotes}
\end{threeparttable}
\end{table*}

Table~\ref{tab:related-positioning} compares LEMUR-Fuzz with four representative works (Semantic-Over-Syntax~\cite{semantics-over-syntax}, Peach~\cite{peach}, ChatAFL~\cite{chatafl}, and MetaMut~\cite{MetaMut}) along five dimensions. For \emph{Key idea}, LEMUR-Fuzz uses semantic-aware mutation to reach deep states and retains AFLNet-style mutation to probe edge cases. This combines protocol-specific mutation with coverage-guided exploration, whereas the representative works focus on semantic-violation testing~\cite{semantics-over-syntax}, type-aware models~\cite{peach}, LLM-generated messages~\cite{chatafl}, or compiler mutator synthesis~\cite{MetaMut}, respectively.

For \emph{Intra-message constraint handling}, LEMUR-Fuzz uses semantic mutators and fixers to generate meaningful field values and repair cross-field dependencies. Semantic-Over-Syntax~\cite{semantics-over-syntax} uses intra-message rules to generate semantic violations; Peach~\cite{peach} depends on custom data models, and ChatAFL~\cite{chatafl} on LLM-guided messages. For \emph{Inter-message constraint handling}, LEMUR-Fuzz uses fixers to repair dependencies between messages. Peach~\cite{peach} requires custom state models, ChatAFL~\cite{chatafl} uses LLM-guided sequences, and Semantic-Over-Syntax~\cite{semantics-over-syntax} does not handle such dependencies. For example, an MQTT sender must not reuse a Packet Identifier while an earlier request using it remains unacknowledged~\cite{oasis-mqtt-v5}; LEMUR-Fuzz extracts this rule and uses a fixer to repair messages that violate it.

For \emph{Validation of LLM outputs}, LEMUR-Fuzz applies metamorphic, parseability, and fixer tests to its generated components before fuzzing. Semantic-Over-Syntax~\cite{semantics-over-syntax} uses schema checks, ChatAFL~\cite{chatafl} relies on repeated runs, and MetaMut~\cite{MetaMut} checks compilation and execution with manual inspection; this dimension does not apply to Peach~\cite{peach}, which has no LLM-generated outputs.

For \emph{Inputs for LLMs}, LEMUR-Fuzz provides protocol specifications to the LLM. Semantic-Over-Syntax additionally uses an ASN.1 (Abstract Syntax Notation One) schema to describe message structures, field types, and basic value and presence rules~\cite{semantics-over-syntax}. LEMUR-Fuzz does not use an ASN.1 schema as LLM input because such a schema alone cannot track cross-message rules, such as the MQTT Packet Identifier constraint above. Instead, the LLM extracts these rules from protocol specifications and generates fixers. This also avoids the added effort of constructing schemas for our ten protocols, whose fuzzed message formats lack official, ready-to-use ASN.1 definitions. Peach~\cite{peach} uses data and state models directly, without an LLM; ChatAFL~\cite{chatafl} provides a protocol name and seeds, and MetaMut~\cite{MetaMut} provides ASTs and mutator templates.

\section{Conclusion}
This paper presents an approach that enhances mutation-based protocol fuzzing with semantic-aware mutators. Built on AFLNet, LEMUR-Fuzz improves branch coverage over state-of-the-art fuzzers and uncovers previously unknown vulnerabilities. Overall, this work demonstrates that LLM-generated semantic-aware mutators are a practical and effective direction for advancing protocol fuzzing.
% \section{Data Availability}
% All artifacts and data required to reproduce our results are available in anonymized form at
% \url{https://github.com/Keyi-Wang/LEMUR-Fuzz}.

\clearpage
\bibliographystyle{ACM-Reference-Format}
\bibliography{mutator/myrefer}

\end{document}